\documentclass{article}
\usepackage{natbib}
\usepackage[french,english]{babel}
\usepackage[utf8]{inputenc}
\usepackage[T1]{fontenc}
\usepackage{times}
\usepackage{graphicx}
\usepackage{tabularx}
\usepackage{enumerate}
\usepackage{url}
\usepackage{booktabs}
\usepackage{amsmath}
\usepackage[usenames,dvipsnames]{color}
\usepackage{amssymb}
\usepackage{multirow}
\usepackage{soul}
\usepackage{authblk}

\newcommand\CO{CO$_\text{2}$}

\newcommand{\itx}[1]{\begin{itemize}#1\end{itemize}}
\newcommand{\x}{\item}

\title{The two sides of the Environmental Kuznets Curve: a socio-semantic analysis}

\date{}

\begin{document}

\author[1]{Telmo Menezes\thanks{menezes@cmb.hu-berlin.de}}
\author[1,2,3]{Antonin Pottier\thanks{pottier@centre-cired.fr}}
\author[1,4]{Camille Roth\thanks{camille.roth@cnrs.fr}}

\affil[1]{Centre Marc Bloch Berlin (CNRS/HU), Friedrichstr. 191, 10117 Berlin, Germany}
\affil[2]{CIRED (Centre international de recherche sur l'environnement et le développement), EHESS, Paris, France}
\affil[3]{Wissenschaftskolleg, Berlin, Germany}
\affil[4]{CAMS (Centre Analyse et Math\'ematique Sociales, UMR 8557 CNRS/EHESS), 54 Bd Raspail, 75006 Paris, France}

\maketitle

\begin{abstract}
Since the 1990s, the Environmental Kuznets Curve (EKC) hypothesis posits an inverted U-shaped relationship between pollutants and economic development. The hypothesis has attracted a lot of research. We provide here a review of more than 2000 articles that have been published on the EKC. We aim at mapping the development of this specialized research, both in term of actors and of content, and to trace the transformation it has undergone from its beginning to the present. To that end, we combine traditional bibliometric analysis and semantic analysis with a novel method, that enables us to recover the type of pollutants that are studied and the empirical claims made on EKC (whether the hypothesis is invalidated or not). We principally exhibit the existence of a few epistemic communities that are related to distinct time periods, topics and, to some extent, proportion of positive results on EKC.
\end{abstract}

\selectlanguage{english} 


\bigskip

The relationship between environmental impacts and economic activity is a hotly debated topic in environmental economics since its inception in the 1960s. At that time, harsh criticisms against economic growth were released as economic expansion was considered the source of environmental degradation. The summit of this wave was the {\em Limits to growth} -- \citet{Meadows1972} report, which is famous for having made the case that economic growth was not sustainable, either because of the lack of raw resources it constantly demands or because of the amount of pollution it unleashes. This call to put growth to an halt raised strong rebuttals among economists who contented that economic progress and substitution can accommodate the economy with any resource scarcity. Fears of expanding pollution were however less easily dismissed. Hence an already famous environmental economist like William Nordhaus could write in 1992 that ``long-run constraints upon economic growth might well exist [...] one possibility is that the
scale of human activity could overwhelm the capacity of the globe to tolerate industrial wastes'' \citep[3-4]{Nordhaus1992}.

Concerns about the need to combine development and environmental protection had led in the meantime to formulating the concept of {\em sustainable development} and making it a goal for the international community. Sustainable development means ``meeting the needs and aspirations of the present without compromising the ability to meet those of the future'', as was stated by the \citet[chap. 1, §49]{Brundtland1987} report, which added that, ``far from requiring the cessation of
economic growth, it recognizes that the problems of poverty and underdevelopment cannot be solved unless we have a new era of growth''. How this new era of growth could be achieved without degrading further the environment was however left unaddressed.

In 1992, the {\em World Development Report} suggested a reassuring way out of the dilemma \citep{WB1992}. It displayed several patterns of environmental indicators depending on country income: monotonically decreasing, monotonically increasing, or increasing then decreasing. According to the latter pattern, the state of the environment, after an initial period of degradation, will automatically improve. The promise was clear: the economic development would somehow take care of its own drawbacks. Environmental degradation would not so much be an issue contrary to accelerating economic growth, as the latter would cure the former. As \citet[491]{Beckerman1992} put it, ``in the longer run, the surest way to improve your environment is to become rich''.

 This inverted-U relationship between economic output and environmental degradation was soon named the environmental Kuznets curve (EKC) hypothesis \citep{Panayotou1993}, from the Kuznets curve, an inverted U-shaped relationship between inequality and economic development that Simon \citet{Kuznets1955}, one of the first national accountants, had observed\footnote{Ironically, this Kuznets curve is no more than a fortuitous relation, see \citet[introduction]{Piketty2014}.}.
According to the EKC hypothesis, environmental degradation has an inverted U-shaped relationship with economic development.
 Many empirical contributions immediately investigated whether such a relationship exists or not, for what type of pollutants, while others pinpointed the problems in estimating its parameters.
  A literature subfield, both theoretical and empirical, grew quickly around the “environmental Kuznets curve”.

Thirty years after these seminal contributions \citep{Grossman1991,Shafik1992,Panayotou1993} more than 2000 articles belong to this subfield. To assess what this literature has finally to say on the validation or not on the EKC hypothesis is challenging, given its size. This is a problem faced by many social scientists: the significant growth in published articles makes it harder to identify, analyze, and assess trends, topics or influential ideas in a given field.

The difficulty of the task motivates the research and the development of an automated analysis, that we present here in this special issue on ``quantitative and computational methods in the social studies of economics''. We hope to foster the application of computational techniques to a broader range of subjects and to stimulate the discussion around their possibilities.

The purpose of this article is first to describe the evolution of the EKC literature. To that end, we quantitatively investigate this literature with tools from social analysis and semantic analysis, hence leading to a socio-semantic analysis. Social analysis refers to the study of networks between actors: here we used the various bibliometric networks that can be derived from the oriented relation that a citation creates between two articles. Semantic analysis refers to the automatic study of the meaning of sentences -- in our case by detecting linguistic patterns that we identified for this specific application -- and it will be applied to the abstracts of the collected articles. 

We aim at assessing how the empirical literature of the EKC has developed and whether it has come to a conclusion. However, such an assessment is more complicated than it may seem. Indeed, the literature on EKC has gone in many different directions and has undertaken significant changes since the first contributions, to the point that validating the EKC hypothesis has no longer a univocal meaning. We therefore also map the development of this specialized research, both in term of actors and of content, and trace the transformation it has undergone from its beginning to the present.

The success of the EKC concept that originates within economics, its spreading in several related disciplines and its evolution offer however broader lessons that go beyond this particular subfield. Our study shows how a concept like the EKC has actually become a keyword on which research from various backgrounds hangs on. This evolution could be seen as another example of economic imperialism but it is actually more accurately described as a combination of attraction of the EKC concept on related disciplines and outpouring of the EKC research in other disciplines. In the process, the EKC research field has lost part of its initial cohesiveness: it has been, in a way, a victim of its own success. 
Our study thus provides with the EKC an interesting case in the social studies of economics to understand the diffusion of economic concepts outside economic literature.

Finally, our article also innovates on the methodological side by combining semantic and socio-bibliometric analysis. Social network analysis has been advocated to be used in the history of economics, for example to study collaboration and interactions in scientific communities \citep{Claveau2018,Herfeld2018}. We use a social network derived from bibliometric data and we supplement bibliometric analysis with an automatic recognition of the claims made by the article. This allows us to recover what an article does or claims to do and thus to map the social relations among actors with the actual content of the scientific disputes between them.
The approach developed in our study can be readily used to investigate other fields or concepts. The case study on the EKC research allows us to reflect on the advantages and limitations of the approach.

The article is organized as follows. Section \ref{sec-literature} reviews the literature on the subfield of the EKC and positions this article among the bibliometric analyses of this subfield. Section \ref{sec-methods} presents the methodology of semantic analysis and the first results in terms of temporal development of the field, type of pollutants that are investigated, and type of claims that are made. Section \ref{sec-network} introduces the analysis of the citation network within the corpus, contrasts its various blocks and the dominant individuals and journals, while discussing its temporal features and relying on the instruments provided by the previous section. Section~\ref{sec-discussion} furthermore discusses limitations of our analysis, followed by a summary of our findings in Section~\ref{sec-conclusion}.

\section{Early history of the Environmental Kuznets curve and literature review}
\label{sec-literature}
The environmental Kuznets curve hypothesis encapsulates the idea that environmental impacts increase then decrease along with economic development. The modern formulation of this idea is the product of several studies conducted simultaneously in the early 1990s. It can be traced back to \citet{Grossman1991}, who studied the environmental impacts of the free-trade agreement between the USA, Canada and Mexico. They observed that the ambient concentrations of some pollutants (SO$_2$ or particulate matter) increase at low income but decrease at high income, hence suggesting that trade liberalization that would foster economic development could also be good for the environment. The same idea also inspired \citet{Shafik1992} who studied more systematically the link between economic growth and environmental quality, considering eight indicators of environmental quality across a large number of countries.  This research was given wide outreach as it was a background article for the {\em World Development Report} 1992 \citep{WB1992}, which incorporated some of its finding as we have explained in the introduction. 

After the publication of the {\em World Development Report}, the number of publications concerned with the inverted U-shaped relationship between the environmental degradation (measured by several indicators of pollutants) and economic development (measured by GDP or GDP per capita) increased quickly and used the phrase ``Environmental Kuznets curve'' coined by \citet{Panayotou1993} at the International Labour Office to refer to this particular relationship. From the start, the EKC hypothesis triggered heated debates. Empirical tests, based on cross-section of countries or panel data, gave varying results, depending on the type of pollutants (water and local air pollution were mostly investigated), the sample of countries or the time period. The actual significance of the EKC was also contested: most articles were interested in the value of the turning point, that is the income after which pollution would decrease, yet \citet{Cropper1994} argued that the height reached by the EKC at its peak was just as important. The political implications of the existence or not of an EKC were no less controversial \citep{Panayotou1997}. Along the lines suggested by \citet{Grossman1991}, the mechanisms that would explain such a relationship were also discussed,  distinguishing the effects of an expanding scale of the economy, of composition change of output, and of the progress of technologies.

In 1997, a special issue in {\em Environment and Development Economics} 
attempted to provide new insights on this growing field. \citet{Barbier1997} in its introduction summed up the main directions taken by the research on EKC.
He synthesized the results obtained for a large varieties of pollutants (classified by broad categories: air pollution, water pollution, deforestation and others) and discuss the level of income at which the turning point occurs. He already pointed to features of the field that will be long lasting, i.e. that the EKC hypothesis remains highly contested and controversial. 

The potential for a proliferation of articles on EKC can already be discerned there.
\citet{Barbier1997} indeed remarked that the EKC is a falsifiable hypothesis and thus lends itself to be tested in various national contexts, with different methodologies, for several pollutants, etc. Because it was formulated in a falsifiable form, yet sufficiently vague (which pollutants? time frame? countries?), one could see in retrospect that it calls for numerous, and possibly endless, empirical studies.

Explaining EKC also raised a theoretical challenge. Some theoretical models were indeed proposed \citep{Andreoni2001,Brock2010,Kijima2010}, but they have not been produced on the same scale as the empirical studies (see \citet{Kijima2010} for a survey).

Surveys were necessary to keep track of the development of the research field and they became more and more dense to address its many branches. \citet{Dinda2004} surveyed the literature, which he already found ``sizeable''. He discussed the empirical estimations and theoretical explanations, as well as the conceptual and methodological critiques raised against the EKC. His survey was 25 pages long and included around 180 references.

The more recent survey by \citet{Shahbaz2019} focused on empirical articles dealing specifically with \CO\ emissions \hbox{i.e.,} almost 200 articles published before 2017. Although they restrict themselves to a single type of pollutant, they find the research to be inconclusive. The results are wildly varying, which they relate, as in the early days of the field, to the many methodologies, contexts, time frames, data sets that the researchers can choose.

Aside surveys directly engaging with the EKC hypothesis and synthetizing results for researchers of the field, several articles examined this field from an external perspective, using bibliometric tools.
\citet{Sarkodie2019} has done a meta-analysis aiming at identifying the value of the turning point found by studies validating the EKC hypothesis (though irrespective of the type of pollutants). They also offer a section with a bibliometric analysis of the EKC research. They mainly investigate author contributions, citations, journals, countries, as well as topics. Their analysis is however static and cannot be used to map the evolution of the field.  \citet{Bashir2021} investigate the citation network of the EKC literature. They compute basic statistics, like the countries or research institutes contributing most to the literature,  the journals in which the literature is published, the most cited studies. They also draw collaboration networks between authors and look for trending topics within the EKC literature.
\citet{Koondhar2021} investigate the co-citation network within the EKC field with the ``CiteSpace'' software. They focus on relations between authors, journals, institutions and keywords.
\citet{Anwar2022} have similarly done a bibliometric review from data of the so-called Web of Science database. They also use the ``CiteSpace'' software to identify key articles according to various provided metrics. However their results are not interpreted: they simply list the articles found and sum them up briefly so that as it stands their work does not allow an understanding of the development of the EKC literature. 
Finally, \citet{Naveed2022} focus on the most influential journals, articles and authors, relying on citations analysis of articles indexed in the ``Scopus'' database. They also analyze and classify the content of the 100 most influential articles.

These studies of varying depth however do not address what we want to investigate. First, the analysis relies on keywords to discuss the content of the article, thus ignoring its meaning, that we can catch with semantic analysis. Second, they are mainly static and they lack an historical perspective, whereas we map the development of the field through time. Third, they do not pay attention to the transfers that occurred across disciplines and journals, which we highlight in our analysis.

\section{Methods and descriptive results}
\label{sec-methods}

\subsection{Corpus building}
We build our corpus from the Scopus database.
Scopus is one of the main bibliographic databases, released by the scientific publisher Elsevier.\footnote{{https://www.elsevier.com/solutions/scopus/how-scopus-works}}
We query Scopus for articles containing the exact phrase ``environmental Kuznets curve'' in the title, abstract or keywords. The search was performed on June 4, 2021 delivering 2709 articles. We collect all possible data that were furnished by Scopus. Here, the analysis will mainly be focused on titles, authors, abstracts and citations. We screen random references to see whether the collected articles were mis-attributed. Given that the syntagm ``Environmental Kuznets curve" is already quite narrow, we find no article that addresses a different topic, so we do not dwell further to exclude wrongly included articles, that is we retain the whole database as the corpus of our analysis.

\begin{figure}[!h] 
\begin{center}
\includegraphics[width=.75\linewidth]{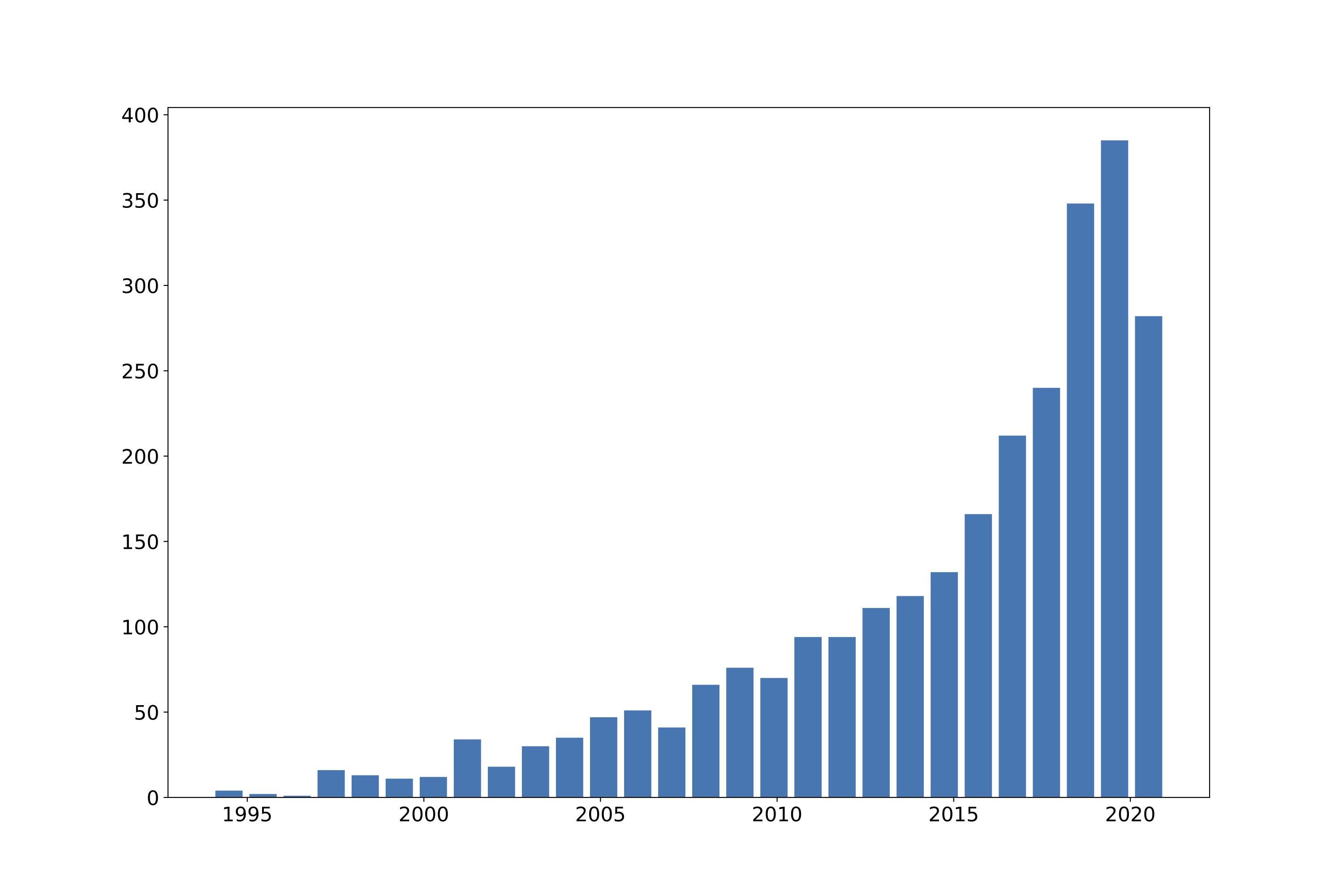}
\caption{Corpus per year of publication}
\label{corpus-temporal}
\end{center}
\end{figure}

Figure \ref{corpus-temporal} shows the number of articles in our corpus per year of publication. It shows a strong increasing trend, as 54\% of the articles have been published in the last four years and a half (since 2017 included), although the first articles are from 1994 (only 2\% have been published before 2000 included). We find that in the last decade the number of article grows at a 20\% rate per annum, far more than the growth rate of between 5\% and 6\% found by \cite{Claveau2016} for articles in economics.

We consider three periods of time: the last two five-year periods (2012-16 and 2017-21), and the beginning of the dataset (1995-2011) gathered in a single period for the sake of significance (there are comparatively very few articles in these early stages, as shown in figure~\ref{corpus-temporal}), to strike a sensible balance between article density and temporal resolution. As a result, the 1995-2011 period comprises 617 articles; 2012-2016, 621 articles, and 2017-2021, 1467 articles. There are 4435 unique authors in the corpus.

Scopus gives for each article of the corpus the list of references of that article. Parsing this list of references, identifying each and finally linking it back to an article of our corpus enables to build the citation network within our corpus. All citation analysis will be based on this network and its derivatives. Hence, all metrics refering to number of citations come from this and so have to be understood as internal \hbox{i.e.,} most-cited articles are articles that are most cited by the articles of the corpus. These pre-processing steps were performed using simple Python scripts that we developed for this purpose.

\subsection{Semantic analysis and semantic hypergraphs}

We develop an automatic pattern recognition approach to extract claims about the EKC from abstracts --- in a nutshell, we are able to classify the recognized claims as positive and negative results on the EKC, and the variables they are based upon. It is especially interesting for an analysis of the field as it allows to reach, to some extent, what an article claims, and in turn to deepen traditional bibliometric analysis by complementing it with semantic analysis. The same abstract may report both positive and negative results. Since our method works at the level of individual claims, we can and do detect such mixed cases. A limitation inherent to the corpus we use is that only abstracts are considered, and sometimes negative results are reported only in the main text of the article.

\newcommand{\Tx}[1]{{\small{\sf #1}}}
\newcommand{\Qx}[1]{\begin{center}\Tx{#1}\end{center}}
\newcommand{\QCx}[1]{\vspace{-.12em}\Qx{#1}\vspace{-.12em}}
\renewcommand{\QCx}[1]{\Qx{#1}}

Our approach is based on \emph{Semantic Hypergraphs (SH)}, a knowledge representation model that is intrinsically recursive and accommodates the natural hierarchical richness of natural language. A full description of SH can be found in \cite{menezes2019semantic}. SH have been recently proposed, but they have already been used for example in the study of the credibility of research impact statements~\citep{bonaccorsi2021credibility}.

One popular contemporary alternative would be to train a classifier with the help of a large language model, e.g. of the BERT family. A disadvantage of this approach is that, given its purely statistic nature, it typically require large training datsets. This disadvantage is accentuated by the fact that this NLP task is somewhat delicate, particularly in identifying if a result is positive or negative, and understanding negation appears to be a typical shortcoming of current deep learning models~\citep{ettinger2020bert}. As will become clear in the subsequent paragraphs, identifying if a result refers to a confirmation or refutation of EKC also requires some nuance. SH makes it possible to infer a useful classifier from a small number of cases -- which is not only important because of the human effort required to annotate training datasets, but perhaps even crucial because our dataset is relatively small.

We give an overview of SH in Appendix~\ref{app-sh}. In a nutshell, SH makes it possible to define and extract relevant linguistic patterns in a semi-supervised manner. More precisely, focusing on a subset of 500 randomly selected sentences, we first extract from the most frequent predicates the ones that are associated with result claims. We add conditions to focus on results related to EKC, including rules to capture the notions of "U-curves" and "N-curves" which, in the corpus, implicitly refer to EKC. We finally take negations into account in order to distinguish between positive and negative results.

On the whole, we design a handful of patterns which are able to classify sentences featuring an EKC claim, and whether it denotes a positive or a negative result.
To evaluate the performance of our semantic classifier, we randomly selected 3 sets of 50 abstracts each, to be manually annotated by the 3 authors, so that each author annotated one of the sets. Notice that abstracts can present no results, only positive results, only negative results, or both positive and negative results. Annotation was performed with no knowledge of the automatic classification. Overall precision and recall~\footnote{\emph{Precision and recall} are common measures in Machine Learning, used to evaluate a classifier's performance. Precision is the fraction of true positives out of all positive predictions, while recall is the fraction of true positives out of all actual positive observations in the data.} results are shown in table~\ref{tab:sh-validation}.

\begin{table}[h!]
\begin{center}
\begin{tabular}{>{\em}l@{\hspace{2em}}cc}
\toprule
EKC validation& \multirow{2}{*}{Precision} & \multirow{2}{*}{Recall} \\
claim type&  &  \\
\midrule
positive result & .809 & .847 \\
negative result & .833 & .366 \\
\bottomrule
\end{tabular}
\caption{Precision and recall of EKC result classifier when compared against a randomly-selected set of 150 manually-annotated articles.}
\label{tab:sh-validation}
\end{center}
\end{table}

We found both precision and recall to be satisfactory for positive claims. For negative claims, precision is satisfactory but recall is poor, which is to say that our classifier markedly underestimates the number of negative claims. We found this to be related to a tendency by authors to present negative results in a less explicit fashion: negative results are often expressed with more convoluted sentences, and with a lot of qualifications. Often, negative results are also implied and thus cannot be identified through direct claims.
A telling example is \citet{Acaravci2010}, a largely cited article (155 times, ranking 7th in the corpus), which is automatically classified as claiming a positive result because the abstract indicates that ``These results support that [sic] the validity of environmental Kuznets curve (EKC) hypothesis in Denmark and Italy''. However, the article investigates the relationship between \CO{} and real GDP per capita for 19 European countries: 
the abstract thus implicitly implies that a relevant EKC relationship was \emph{not found} for the remaining 17 countries, which we confirmed by examining the article.

The much larger recall error of the algorithm for negative results leads us to correct all values found by the algorithm by multiplying raw quantities of results by the associated precision and dividing them by the associated recall.\footnote{Precision corresponds to the proportion of correctly detected items of a certain category among those detected as belonging to that category, while recall is the proportion correctly detected items among those belonging to that category. Hence, precision/recall is the rate of correctly detected items per items detected as belonging to a category \citep[for more context on such correction factor see e.g.][]{schmid2020prey}.} For instance, if a negative result is detected in $N$ articles, we consider that there are actually $\frac{N\cdot 0.833}{0.366}$ such articles. Notice that negative claim precision also contributes to positive claim precision: even if negative recall is relatively weak, negative precision is still quite useful because it prevents claims from being wrongly classified as positive claims.

\bigskip

\begin{figure}[!t] 
\begin{center}
\includegraphics[width=.5\linewidth]{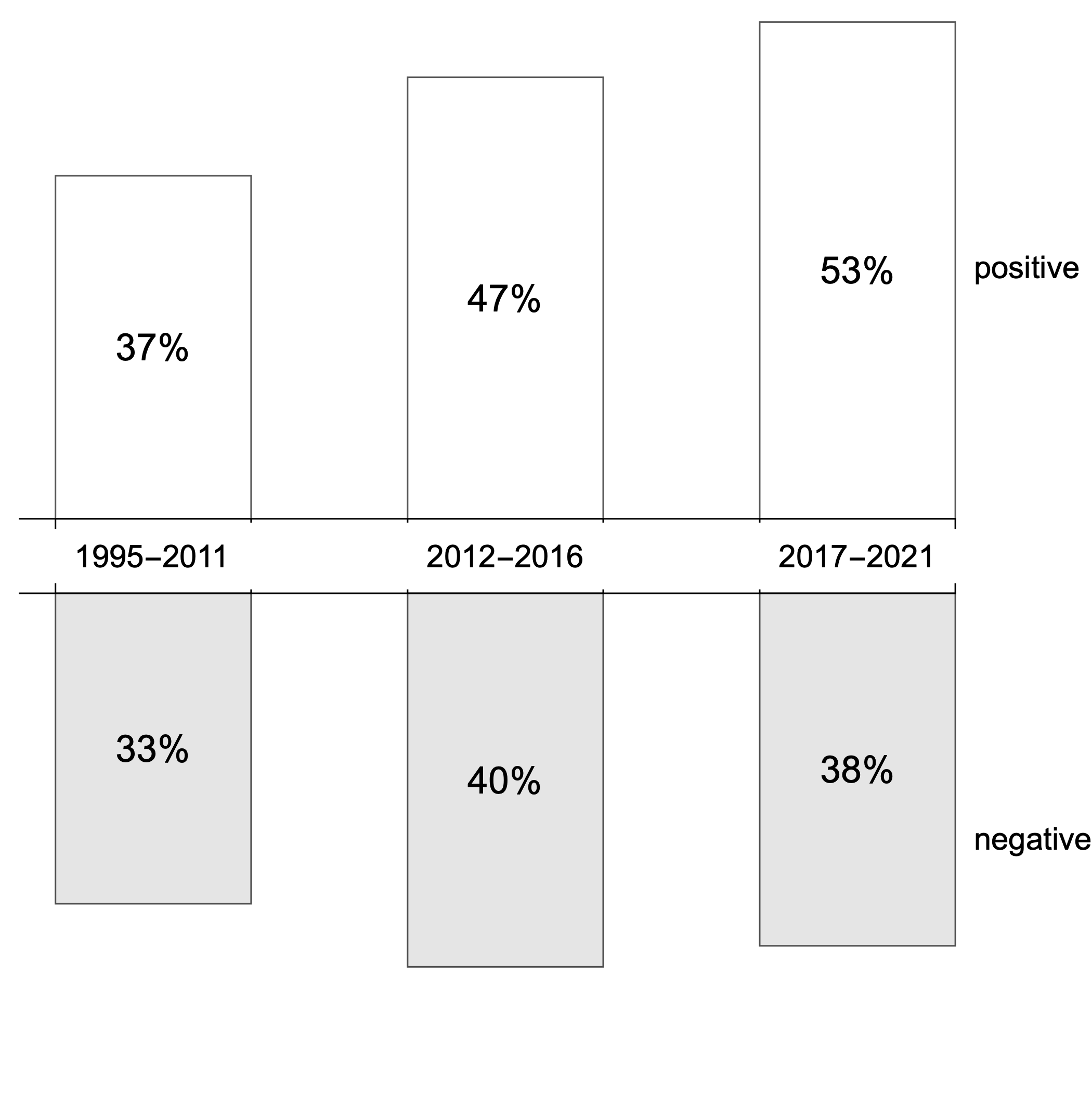}
\vspace{-1em}
\caption{Estimated percentage of articles with a positive or with a negative result.}
\label{fig:articles-posneg}
\end{center}
\end{figure}

In figure~\ref{fig:articles-posneg} we plot the ratio of articles with at least one positive result and the ratio of those with at least one negative result, knowing that a small portion of articles feature both.
We see that the proportion of articles with positive results and that of articles with negative results do not exhibit a very marked difference. Even though there has been a modest albeit increasing lead in the number of articles with positive findings over the recent years, there nonetheless remains a substantial proportion of articles with negative results. From this analysis of the meaning of the abstracts collected in our corpus, we can confirm that the literature on EKC is generally unconclusive, as stated by several observers or actors of the field \citep{Shahbaz2019,Haberl2020}. In other words, the evidence cannot be said to be settled and there is an on-going controversy, which may reflect different scientific practices and epistemic communities. We will comment more on that in section \ref{sec-diverging}.

\subsection{Topics}
The EKC is a generic concept that can be applied to the relationship between an economic development variable (such as GDP) and various pollutants and environmental pressures.  We now identify which kind of environmental variables have been investigated in a given article.

Once again we take advantage of SH representation. We found that the pattern ``relationship between X and Y'' is prevalent across the corpus, and that it largely corresponds to a relationship between an economic variable and an environmental variable.\footnote{The pattern can be represented in SH as:
\Tx{\scriptsize(between/B relationship/C (and/J X Y))}.}
We exploited it to identify all the environmental variables present in X or Y. The most common ones were manually grouped into eight categories: greenhouse gases (labeled as ``GHG'', which includes for a very large part \CO-related research), energy, water, waste, environmental footprint, sulphur oxides (SOx), nitrogen oxides (NOx) and local air pollutants (like particulate matter or carbon monoxide, this category does not include SOx nor NOx). We treat these environmental variable categories as topics.

\begin{figure}[!t] 
\begin{center}
\includegraphics[width=.6\linewidth]{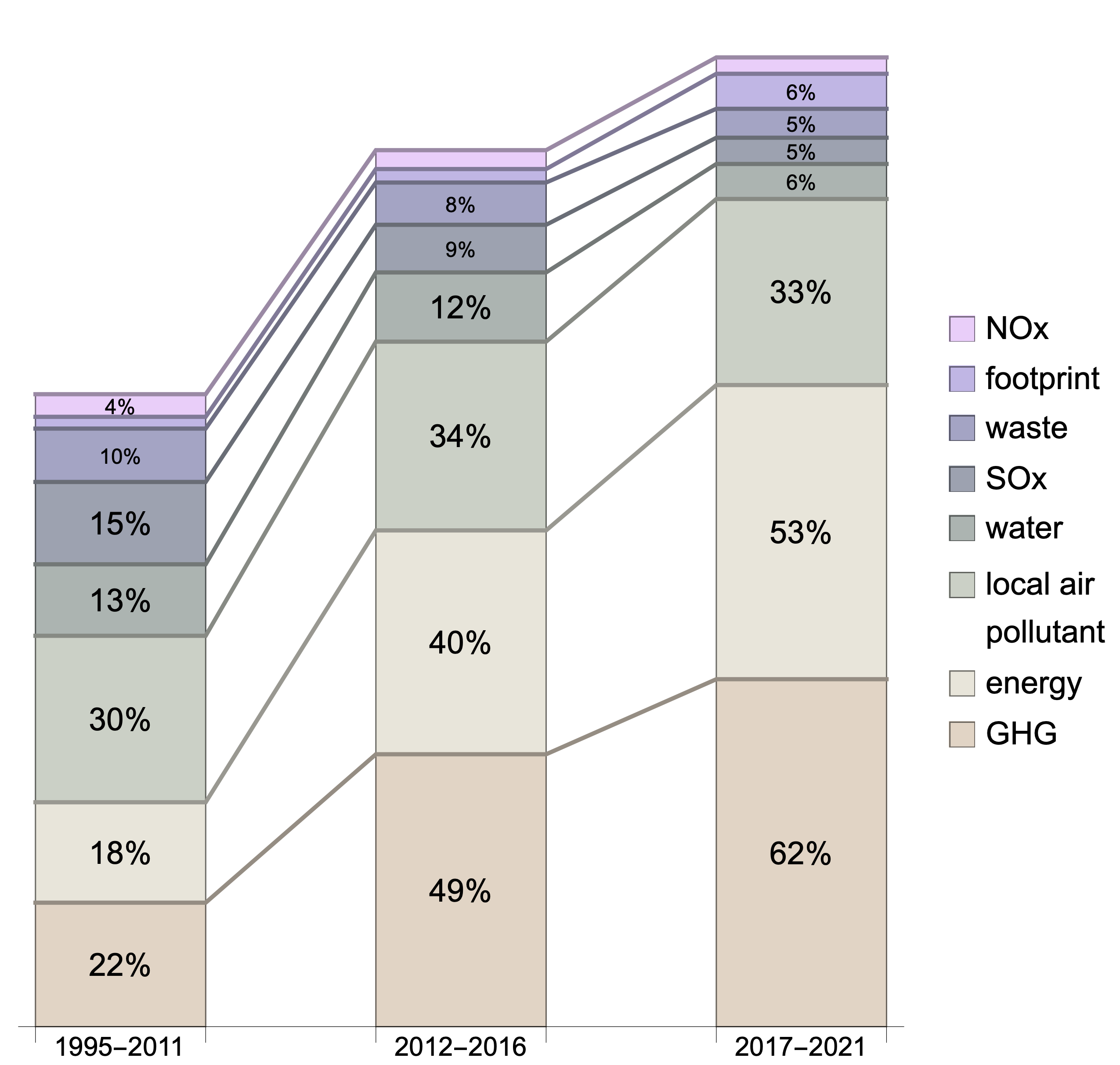}
\caption{Percentage of articles evoking a relationship on a given topic (sums over 100\% as an article may address several topics, bar elements representing less than 4\% were not labeled).}
\label{fig:articles-topic}
\end{center}
\end{figure}

On the whole, 82\% of articles were found to contain at least one of these environmental variable categories. On figure~\ref{fig:articles-topic} we present the ratios of articles mentioning each category. Ratios sum to more than 100\% since, although topics are non overlapping, an article can be classified in several categories: an article can analyze the EKC both for air and water pollutants. Energy and its various vectors (oil, coal, gas) were frequently associated with income, yet it appears that articles are often both categorized as energy and carbon. This is because the articles that investigated the link between GHG emissions and income often add various form of energy as control variables.

This analysis yields three main results. First, the strongly increasing focus on GHG/\CO{} emissions and energy, which was also found by \cite{Sarkodie2019} using a different method (keyword analysis). This focus contrasts with what one might have anticipated based on the early developments in the field. Indeed, although \citet{Shafik1992} considered \CO{} within their set of pollutants, the early decade of research on EKC was not strongly concerned with \CO. \citet{Dinda2004}'s review mentions research on \CO{} but focuses on local air pollution and water pollution. One could therefore have expected that the literature would go on focusing on these pollutants, as the EKC hypothesis was heavily debated for these ones (see for example \citet{Stern2001} on sulfur emissions). Furthermore, as \citet[p.~441,~449]{Dinda2004} makes clear, the literature at that time did not find the evidence of EKC for global pollutants like \CO. On the contrary, from 2014, the articles identified as investigating \CO{} or GHG constitute more than half of our corpus, representing above 55\% thereof in recent years. The trend is quite massive and shows no sign of losing steam. Second, while seminal publications discussed various pollutants, especially air pollutants such as nitrogen and sulfur oxides (NOx and SOx), and various environmental stressors such as waste, more recent publications do not seem to focus explicitly on such specific pollutants: the shares of SOx, NOx, waste as well as water [pollution] strongly decreased; only the share of the generic reference to ``local air pollutants'' remained stable.
To summarize, we witness a strong turn to energy and GHG-related pollutants from around 2012.

This turn can be explained. First and foremost, the growing number of articles targeting GHG emissions has to be related to the increased salience of climate change as the environmental problem vital for human societies. This means that researchers are more likely to focus on GHG and neglect other environmental problems, but also that research is more easy on this topic because of the large availability of dedicated databases (on GHG emissions) and of the ease of attracting fundings. Furthermore, methodological changes are also involved in this turn.
\citet{Ang2007} was the first to bring together a literature focused on the relationship between economic output and energy, and the EKC literature, hitherto focused on the relationship between economic output and pollutants (here specifically \CO\ emissions). This promoted the integration into the EKC literature of studies considering both energy and GHG emissions, while transforming the EKC itself, as it became a relationship between economic output and GHG emissions within a broader framework.
We come back to this transformation in the discussion (see section \ref{sec-discussion}).

A cross analysis of positive/negative results \hbox{vs.} topics makes it possible to refine the picture further. Table~\ref{tab:fig-pos-neg-co2} provides the relative proportion of positive \hbox{vs.} negative results for each topic through time --- in other words, it paints the categories of figure~\ref{fig:articles-topic} with the categories of figure~\ref{fig:articles-posneg}. We see that the literature of the last period 2017-21 features a majority of positive results (proportion $>1$) for all topics. This reflects the aggregate trend previously observed for that period, for which the share of positive results significantly increases (to reach 53\% vs 38\%). This trend is however not uniform: results related to footprint, GHG, local air pollutants, NOx and, to a smaller extent, SOx, experience the strongest increase in positive results. More interestingly, for half of the categories (local air pollutants, footprint, water and NOx), trends are reverted from a majority of negative results to a minority; the same could almost be said for GHG (from half-half to strongly positive-leaning results). Waste is the only topic that exhibits a decrease in the proportion of positive results from the first to the final period. Thus, while the increase in topic-related research suggests that the field is getting more specialized or, at least, giving more attention to specific relationships, this tendency also comes with an evolution of the imbalance around the confirmation or refutation of EKC that seems to affect more certain topics than others. We shall see below how this topical specialization is distributed on the author network.
\newcommand{\mx}[1]{\multirow{2}{*}{#1}}
\begin{table}\setlength{\tabcolsep}{.47em}
\centering\begin{tabular}{>{\em}c@{\hspace{2.em}}cccccccc}
\toprule
&\multicolumn{8}{c}{\bf\normalsize Topic}\\
&\mx{GHG} &\mx{Energy} &Local air &\mx{water}&\mx{SOx}&\mx{Waste}&\mx{Footprint}&\mx{NOx}\\
\normalsize Period	&	  &  	   &pollutants\\
\midrule
1995-2011&1.03&1.29&0.84&0.81&1.18&{\bf1.10}&0.31&0.91\\
2012-2016&1.05&1.09&1.15&{\bf 1.03}&1.00&0.67&0.42&0.92\\
2017-2021&{\bf1.55}&{\bf1.39}&{\bf1.64}&1.01&{\bf 1.39}&1.01&{\bf 1.07}&{\bf 1.50}\\
\bottomrule
aggregate&\em 1.35&\em 1.31&\em 1.32&\em 0.95&\em 1.19&\em 0.94&\em 0.88&\em 1.14
\end{tabular}
\caption{\label{tab:fig-pos-neg-co2}Evolution of the ratio of articles with positive \hbox{vs.} negative results, broken down by topic, computed over the three defined time periods (figures in bold indicate for each topic the period of maximum ratio) and for all periods (``aggregate'').}
\end{table}

\section{Citation network and blocks}
\label{sec-network}

To have a better sense of how the field has developed and why, we now turn to the analysis of the author citation network.

To begin, we compute for each author the number of times their articles have been cited. As said before, we only consider citations internal to the corpus. This means that the citations attracted reflect the influence of the articles in the field of the EKC research. With this, we cannot assess the impact of the articles of the corpus in the environmental economics literature in general, nor are we able to assess to the influence of articles outside the corpus on the corpus.

We concentrate on the most (internally) cited authors, listed in table \ref{most-cited-authors}, for whom we can already distinguish several patterns. I.~Öztürk and M.~Shahbaz entered the fields in the 2010s, and since then they have published a large number of articles. D.I.~Stern has been present in the field over the whole time span, with a large yet lower number of articles per year. Soumyananda Dinda has been mainly influential through his review of 2004 published in {\em Ecological Economics} (673 out of 879 citations).

\begin{table}[h!]
\begin{center}\setlength{\tabcolsep}{1em}
\begin{tabular}{lrrr}
\toprule
Author &\#A & \multicolumn{1}{c}{$c$} &\multicolumn{1}{c}{Activity} \\
&&&\multicolumn{1}{c}{time span}\\\midrule
Öztürk I.	    & 35	& 1779  & 2010-2021 \\
Stern D.I.	    & 17	& 1622	& 1996-2020 \\
Shahbaz M.	    & 49	& 1524	& 2012-2021 \\
Dinda S.	    & 7	    & 879	& 2000-2009 \\
Al-Mulali U.	& 13	& 756	& 2015-2019 \\
\bottomrule
\end{tabular}
\caption{Five most cited authors. \#A is the number of articles, $c$ number of citations.}
\label{most-cited-authors}
\end{center}
\end{table}

The table is a little bit different from the one obtained by \citet{Sarkodie2019}. First two authors are the same, although in reverse order. Shahbaz and Dinda are also present in the top-five  most cited authors of \citet{Sarkodie2019}. The differences seem to stem from three facts. First \citet{Sarkodie2019} use the \emph{Web of Science} database to extract their corpus whereas we rely on \emph{Scopus}. Second, they use metadata citation counts that apply to the whole Web of Science database and thus are likely to include citations outside of the EKC research field. Third, our analysis is done three years later. As Öztürk is a prolific and active author (as we will see), this may explain why he has overcome Stern in the mean time. 

\subsection{Blocks of authors}
We now look into more structural characteristics of the citation network. To do so, we extracted the citation network between authors from the corpus from the citations between articles. Citations are matched to articles in the corpus if both their titles (case-insensitive) and years of publication are the same. Each citing article induces edges of weight 1 from all its authors to each author of the cited article. This is a directed weighted network, with each edge weight indicating the number of times the origin cites the target. This network is very dense, so to make it tractable we adopt a simple edge removal and graph simplification strategy.\footnote{A myriad of graph simplification methods (variously denoted as graph sampling, reduction, compression) have been proposed by the network science community over the last two decades, principally aimed at easing structure visualization or statistical feature computation \citep{zhao2020preserving}. Node-centric edge sampling \citep[such as favoring high-degree nodes e.g.,][]{ribeiro2010estimating} and network-level edge selection \citep[such as minimal spanning trees e.g.,][]{van2008centrality} are typical strategies fulfilling relatively distinct goals: while the former seem to operate on local structure, the latter rather help in preserving a sense of global connectivity \citep{jalali2016social}, even though they can do both \citep[respectively][]{leskovec2006sampling,jia2008visualization}. As we are interested in blocks we rather focus on the former and adopt a very simple approach of uniformly pruning outgoing edges in order to concentrate on the most important targets, while preserving the structure of incoming edges i.e., received authority from peers -- admittedly a key feature of the citation networks we study. While much simpler, it is close in spirit to node-centric sampling methods which tends to favor links to higher degree nodes and demonstrate satisfying preservation of community structure \citep{maiya2011benefits,ahmed-2011-network,voudigari2016rank}.}
We consider only 3 outgoing edges per author, pointing at the 3 most cited target authors by each actor of origin. This means that every node has out-degree 3 and any number of incoming edges (in-degree). Furthermore, we consider only authors that received at least 10 citations. We have found these thresholds to be remarkably robust. For example, employing a threshold an order of magnitude higher (top 30 most cited targets) produces very similar block partitions. To also be able to observe the temporal evolution of the network, we compute cumulative networks for three time periods starting in 1995 and ending, respectively, in 2011, 2016 and 2021, the last one corresponding to the full dataset.

\begin{figure}[th!]\centering
\includegraphics[width=.84\linewidth]{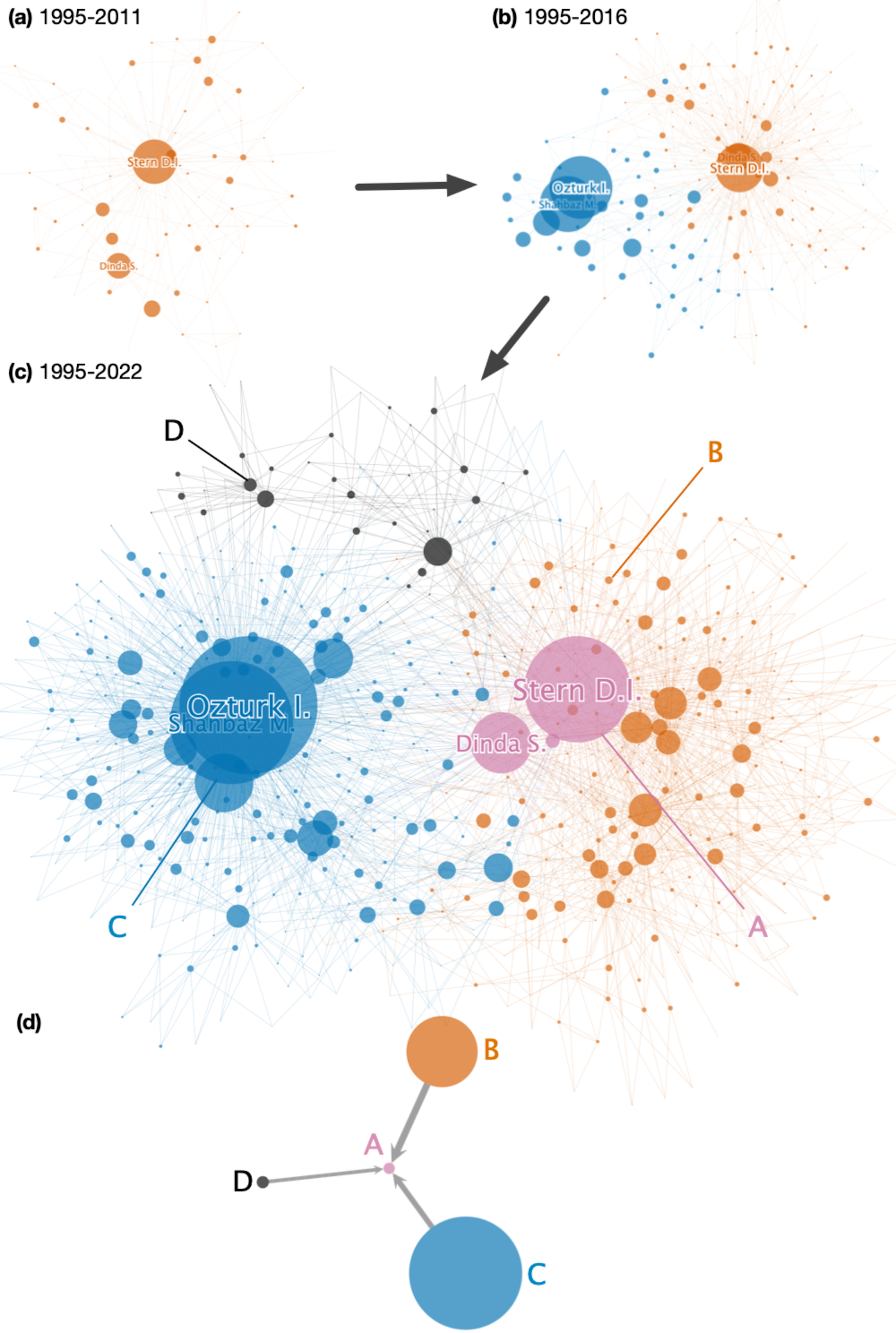}
\vspace{-1em}
\caption{
Graphs of citations from authors to authors (or blocks of authors to blocks of authors in the last panel), colored by blocks determined by SBM. Node radius is proportional to in-degree. Three periods (cumulative): (a) 1995-2011 (b) 1995-2016 and (c) 1995-2022. (d) Graph of author blocks for the last period. Edge thickness is proportional to the overall probability $p$ of any given author from one block citing any given author in the other block. Very low probabilities of connection ($p \le 0.01$) are not depicted. Node radius is proportional to the total number of citations (for the entire corpus) received by the authors in the block.}
\label{fig:citation-graphs}
\end{figure}

We then use Stochastic Block Models (SBM)~\citep{holland1983stochastic, peixoto2018nonparametric} and the open-source library \emph{graph-tool} to infer the structure of the network~\footnote{https://graph-tool.skewed.de}. SBM is a generative model, where each node is assigned to a given block. SBM blocks are such that nodes of a given block connect to nodes of another block (that can be the same) with a constant probability. SBM jointly infers block membership and the matrix of connection probabilities, such that the likelihood of the observed network given the generative model is maximized. Put differently, blocks represent similar connection patterns from the nodes of that block towards nodes from other blocks. One feature of SBM is that maximizing this likelihood also implies a minimization of the amount of information necessary to describe the model (description length). In other words, there is a preference akin to \emph{Occam's Razor} for the simpler model. We performed model selection based on description length and opted for \emph{degree-corrected} SBM~\citep{karrer2011stochastic}, since it achieved significantly lower description lengths in comparison to standard SBM. Given that SBM inference is a stochastic non-deterministic process, we performed it 10K times and selected the one that achieved lowest description length. We arrived at a partition in 4 blocks for the full network covering the entire time period under study, as depicted in figure~\ref{fig:citation-graphs}(c).

Blocks are distinguished by significant linking behavior differences, both inwards and outwards, which can be interpreted as an indication of their role in the network.
For example, block A contains a small number of highly cited authors. A tends to be cited by all blocks and to not cite any other blocks, while all others tend to cite A.

Figure~\ref{fig:citation-graphs}(c) shows us that two large blocks (B and C), of comparable size, account for the majority of authors in the network. Given that the two large blocks are not connected in figure~\ref{fig:citation-graphs}(d), this could suggest the existence of two separate factions with the similar internal dynamics of a large ``following'' and a smaller number of highly influential authors. Notably, Stern belongs to block A and Öztürk to block C. While block A is clearly influential to all the other blocks, its topological insertion suggests its closer alignment with block B. The remaining block (D) is dominated by Chinese authors publishing articles with a strong focus on China.

Notice that figure~\ref{fig:citation-graphs}(d) is rendered with a threshold of probability of connection of $p > 0.01$, and that there is some probability of connection between all pairs of blocks that is greater than zero, and that such connections do exist. Furthermore, the formation of such connections is affected by size effects, which is why even with low affinity below the 0.01 threshold, blocks B and C still have a good amount of connections between them. It is also affected by age: an older article had more time to be cited and can only cite articles that were published before it.

In figure~\ref{fig:citation-graphs}(d) we can observe another interesting topological fact: although Stern has been overtaken by Öztürk in terms of number of citations received within the EKC literature, the small block of authors that he belongs to appears to remain the most central and influential. We can see that all other blocks have a strong tendency to cite A, and that A is in fact the only block \emph{for which this is the case}. Another indication of this asymmetry is the more casual observation that Öztürk cites Stern's articles thirteen times, whereas Stern does not cite Öztürk at all -- this being true also for the full citation graph, not just the simplified one with the 3 outgoing edges per author. When interpreting this asymmetry, it is also good to keep in mind that Stern's work is older than Öztürk's.

In table~\ref{tab:citation-blocks} we present a set of metrics for the blocks that are not of a topological nature. This means that the distinctions that these metrics provide are not implied by the network structure, and therefore help to strengthen the hypothesis that these blocks do indeed correspond to different practices within EKC research. Blocks A and then B have a lower ratio of positive results than C. The ratio of negative results is more similar across blocks, except for A. There is also a clear difference in endogamy, with  C being the most endogamic and both A and B the least. Interestingly, block D appears to be a middle ground between A+B and C in all metrics, as well as in its topological insertion in the citation network. This highly regional block might be influenced by the two main blocks.

\begin{table}
\begin{center}
\begin{tabular}{>{\em}crrrccccr}
\toprule
Block & \#Authors & Pos & Neg & Endogamy & {Year} & \#A & $\overline{\text{\#A}}$ & $k$\\
&&&&&&\\\midrule
A & 3 & 0.354 & 0.084 & 0.003 & 2007.9 & 9.00 & 0.79 & 5.0 \\
B & 360 & 0.422 & 0.416 & 0.033 & 2012.1 & 3.11 & 0.76 & 4.9 \\
C & 406 & 0.562 & 0.407 & 0.102 & 2017.9 & 4.47 & 1.24 & 9.2 \\
D & 57 & 0.492 & 0.441 & 0.085 & 2016.8 & 5.96 & 0.96 & 15.7 \\
\bottomrule
\end{tabular}
\caption{Various metrics per author block, including ratios of positive and negative EKC claims, endogamy measured as ratio of citations received from co-authors, mean year of publication, mean number of articles per author (\#A) and further normalized per year ($\overline{\text{\#A}}$), and mean number of unique co-authors ($k$).}
\label{tab:citation-blocks}
\end{center}
\end{table}

The temporal aspect is also of interest. Considering the mean year of publication for the articles by authors in each block, we can see that block A is the oldest, followed by B, and then C and D are much more recent. This invites the hypothesis of a shift in practices having taken place during some period in time. In fact, the three steps of the longitudinal study, corresponding to the panels (a), (b) and (c) of figure~\ref{fig:citation-graphs}, reveal a complexification of the citation network from uni- to bi-polarity. Blocks in the two first networks were colored according to the block in the final network with which they have the highest Jaccard similarity. It is clear that the field was dominated by Stern and Dinda in the first period, then assigned to block B. Öztürk and Shahbaz appear in the second period along with block C, with the bipolarity already fully developed, and Stern still appearing as the most influential member of block B. In the final period, Stern and two other influential and seminal authors detach into block A and block D is formed.

Finally, the breakdown of the presence of each topic for each block, shown in table~\ref{tab:topics-blocks}, also paints a quite heterogeneous picture. It indicates a certain level of focus and specialization proper to some blocks: for instance, A on SOx, C on GHG and energy and D on local air pollutants.

\begin{table}[t]
\begin{center}
\begin{tabular}{>{\em}ccccccccc}
\toprule
Block & GHG & Energy & Local air & Water & SOx & Waste & Footprint & NOx \\
 &  &  & pollutants &  &  &  &  &  \\
\midrule
A & .333 & .074 & .222 & .000 & {\bf.333} & .000 & .000 & .000 \\
B & .345 & .257 & .344 & .123 & .109 & .084 & .020 & {\bf.042} \\
C & {\bf.664} & {\bf.618} & .310 & .062 & .039 & .043 & {\bf.085} & .023 \\
D & .436 & .392 & {\bf.537} & {\bf.128} & .084 & {\bf.106} & .018 & .040 \\
\bottomrule
\end{tabular}
\caption{Percentage of articles mentioning each topic for each author block. Bold figures indicate the block where a topic has the highest presence.}
\label{tab:topics-blocks}
\end{center}
\end{table}

\subsection{Focusing on the two leaders}
David I. Stern, the second most cited author, has been active in the field for 25 years. He began his career\footnote{Information available at http://sterndavidi.com} with a PhD in Geography from Boston University. From 1996, he was a Research Fellow at the Centre for Resource and Environmental Studies in Australia. He is currently Professor at the Crawford School of Public Policy, Australia. He was associate editor of {\em Ecological Economics} (which appears in Table~\ref{most-publishing-journals}) from 2002 to 2018 and belongs to its editorial board since then. His most cited article is ``Is There an Environmental Kuznets Curve for Sulfur?'' \citep{Stern2001} which uses panel data to investigate the EKC for sulfur emissions and essentially shows that earlier findings of an EKC can be explained by a sample restricted to high-income countries: an EKC could not be found using the global sample.

Ilhan Öztürk, the first most cited author according to our investigation, is a newer scholar who completed his PhD in Economics in 2009 from \c{C}ukurova University in Turkey and began publishing on EKC in 2010.\footnote{Information available at {https://www.cag.edu.tr/en/academic-staff/104/about}} He has been working at \c{C}a\u{g} University since 2000, where he became a professor in 2017.  Öztürk is editor-in-chief of the {\em International Journal of Energy Economics and Policy} (which appears in Table~\ref{most-publishing-journals}), founded in 2011 and edited by EconJournals, a platform run by Ilhan Öztürk. This publisher has two other journals: Ilhan Öztürk is also editor-in-chief of the first and co-editor of the second.
On EKC, his most cited article is  ``Investigating the environmental Kuznets curve hypothesis in Vietnam'' \citep{Almulali2015}. It uses time-series of carbon emissions, GDP, and several controls (imports, exports, various forms of energy) to test for the EKC and finally rejects it.

Öztürk and Stern are the two most cited authors but, given the numbers available in table \ref{most-cited-authors}, they seem to be inserted differently in scientific networks.
As said before, Stern has published less than Öztürk in our corpus, 17 articles against 35, but within 25 years of activity compared to 12, which corresponds respectively to 0.68 articles on EKC per year \hbox{vs.} 3.2. This difference is also visible in the number of unique co-authors: only 10 for Stern \hbox{vs.} 83 for Öztürk. 
These different practices extend to citations, whereby the 1808 citations of Öztürk are more concentrated than the 1644 citations of Stern, in terms of both citing authors and citing articles: 1551 distinct authors cite Öztürk \hbox{vs.} 1992 for Stern (hence 1.17 cites per author compared with 0.83), and 794 distinct articles for Öztürk \hbox{vs.} 1042 for Stern (2.28 cites per citing article \hbox{vs.} 1.58).
We can also note that 11\% of articles citing Öztürk are from him and co-authors, whereas only 0.6\% for Stern.
One finds similar patterns when extending the analysis to the most cited authors: Shahbaz (2.28 cites per citing article, 1.18 cites per citing author) is very close to Öztürk, whereas Dinda is close to Stern (1.13 cites per citing article, 0.56 cites per citing author). 

All this points to distinct practices in scientific writing and publishing. Statistics of table~\ref{tab:citation-blocks} confirm this trend on a broader scale. Block A presents the highest number of articles per author, while B and C have lower similar values at close to half that of A. Yet, blocks A and B correspond to authors with a longer activity span and who started publishing earlier in the field. The average number of yearly articles shows that, on one side, blocks A and B (around Stern) are close to twice less prolific than block C (D representing again a middle ground). Just as we observed on Stern \hbox{vs.} Öztürk, blocks A and B have a lower number of unique co-authors than block C and, noticeably, D. In a nutshell, the block around Öztürk exhibits distinct publication behaviors: its authors arrived later, yet they publish more often, which contributes to inflating the number of articles, and with more distinct people.

\subsection{A shift in journals}

We observe a similar divergence when looking at publication outlets.
Overall, we found 733 unique outlets in our corpus, with a mean of 3.7 articles per outlet. As is common, we have a large dispersion and a long tail of 642 outlets with no more than 4 articles, representing 35\% of the corpus. On the contrary, there are 18 outlets that have published at least 20 articles on the EKC, and these 18 journals represent together 42\% of our corpus.
We list these journals in table~\ref{most-publishing-journals}, along with their temporal profiles which describe for each journal the proportion of EKC articles that fall in each of the three time periods. 
\newcommand{\zerofiveinx}[1]{\parbox[c]{1em}{\includegraphics[width=0.45in]{#1}}}
\begin{table}[h!]
\begin{center}
\begin{tabularx}{\linewidth}{>{\footnotesize}l>{\em}Xrp{.75cm}cc>{\em}c}
\toprule
&\multirow{2}{*}{Journal}& \multirow{2}{*}{\#A} &\multicolumn{1}{c}{Temporal} & \multicolumn{2}{c}{\small\% Articles citing} & \multirow{2}{*}{$r$} \\
&&&\multicolumn{1}{c}{profile}&Stern &Öztürk&\\
\midrule
1989&Ecol. Econ. & 118 &\zerofiveinx{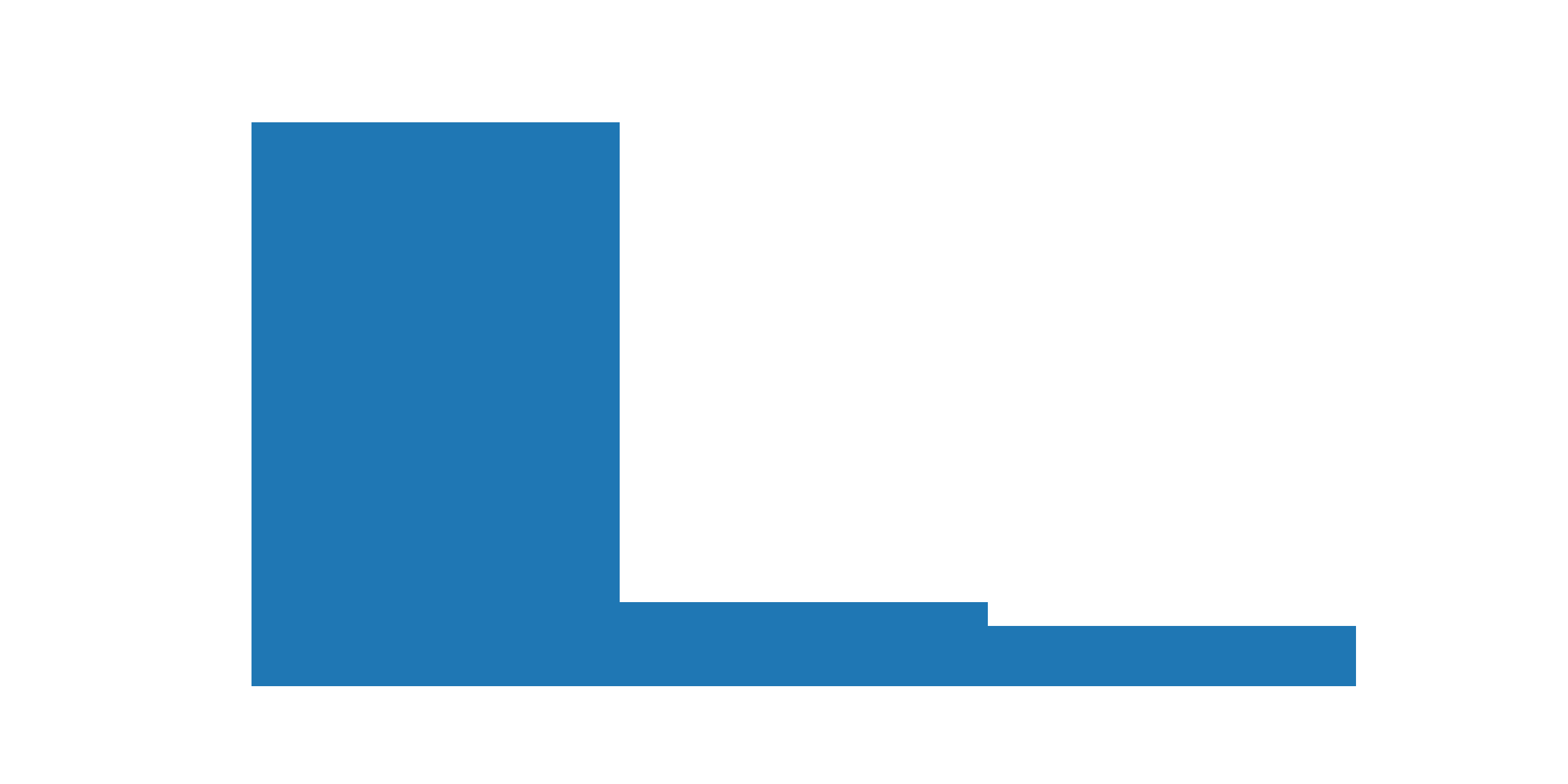} & 0.63 & 0.02 & 0.97\\
1991&Environ. Resour. Econ. & 38 & \zerofiveinx{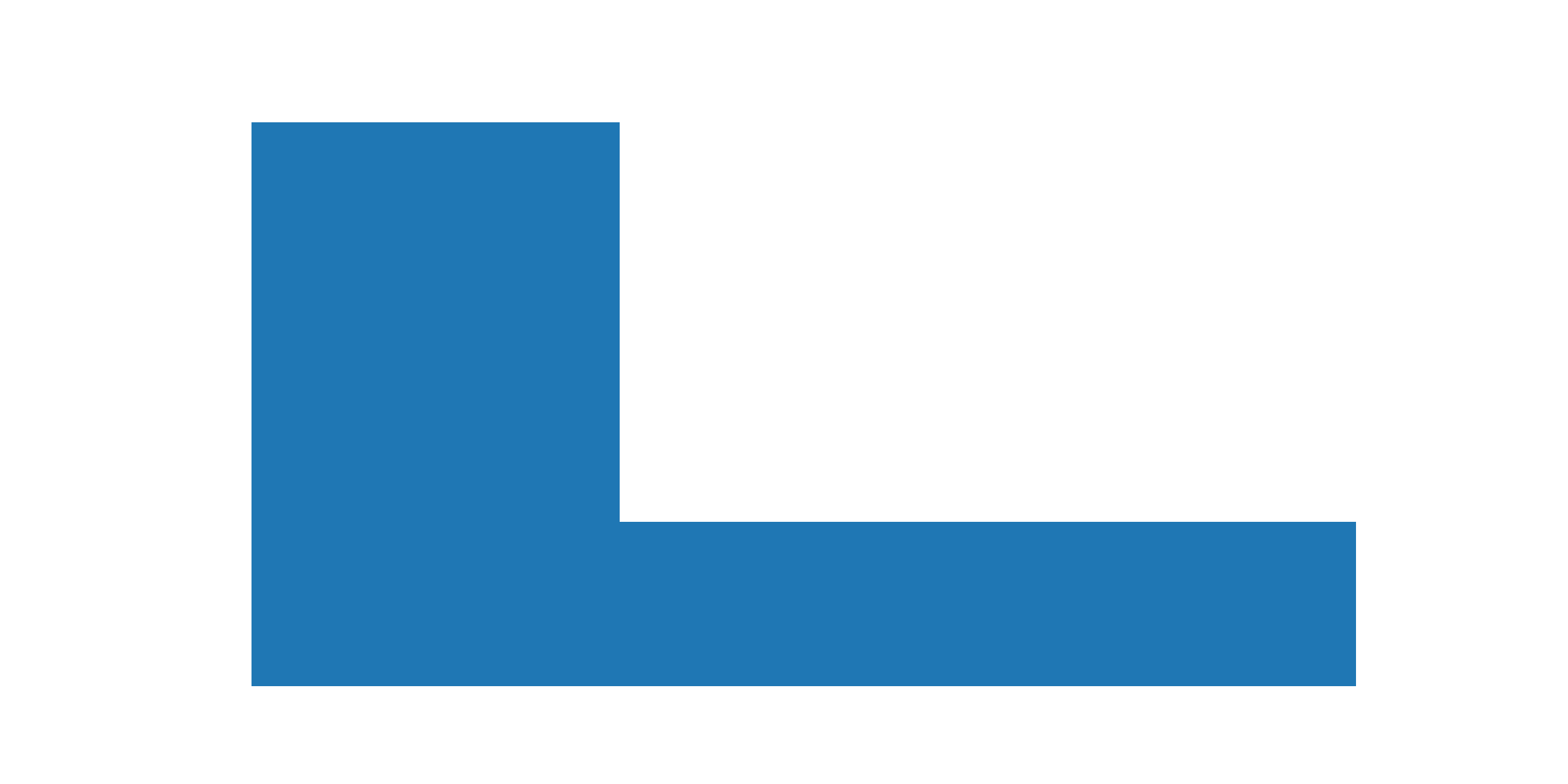}& 0.50 & 0.03 & 0.95 \\  
1996&Environ. Dev. Econ. & 36 & \zerofiveinx{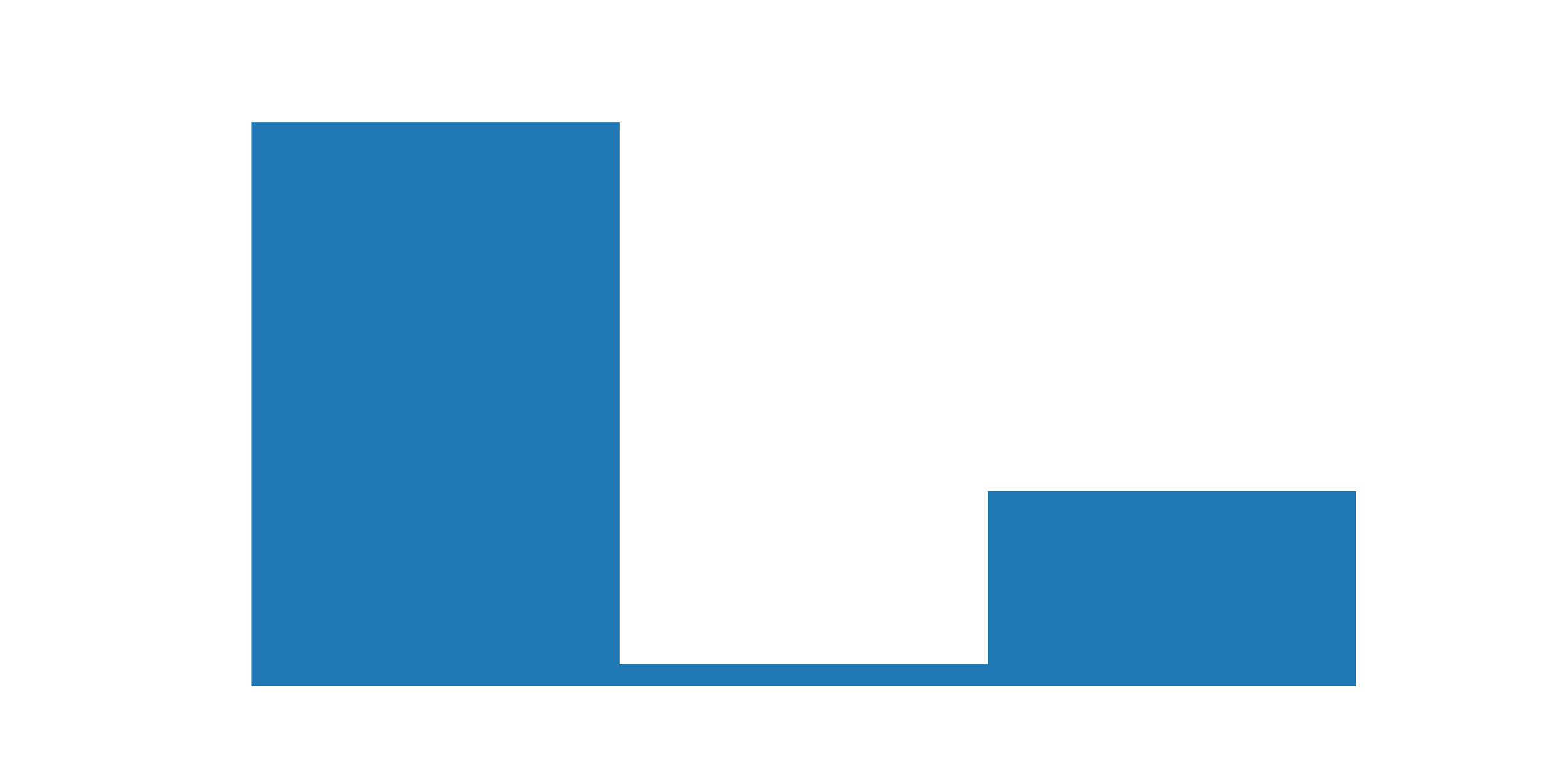} &0.56 & 0.03 & 0.95\\
2001&Int. J. Global Environ. Iss. & 24 & \zerofiveinx{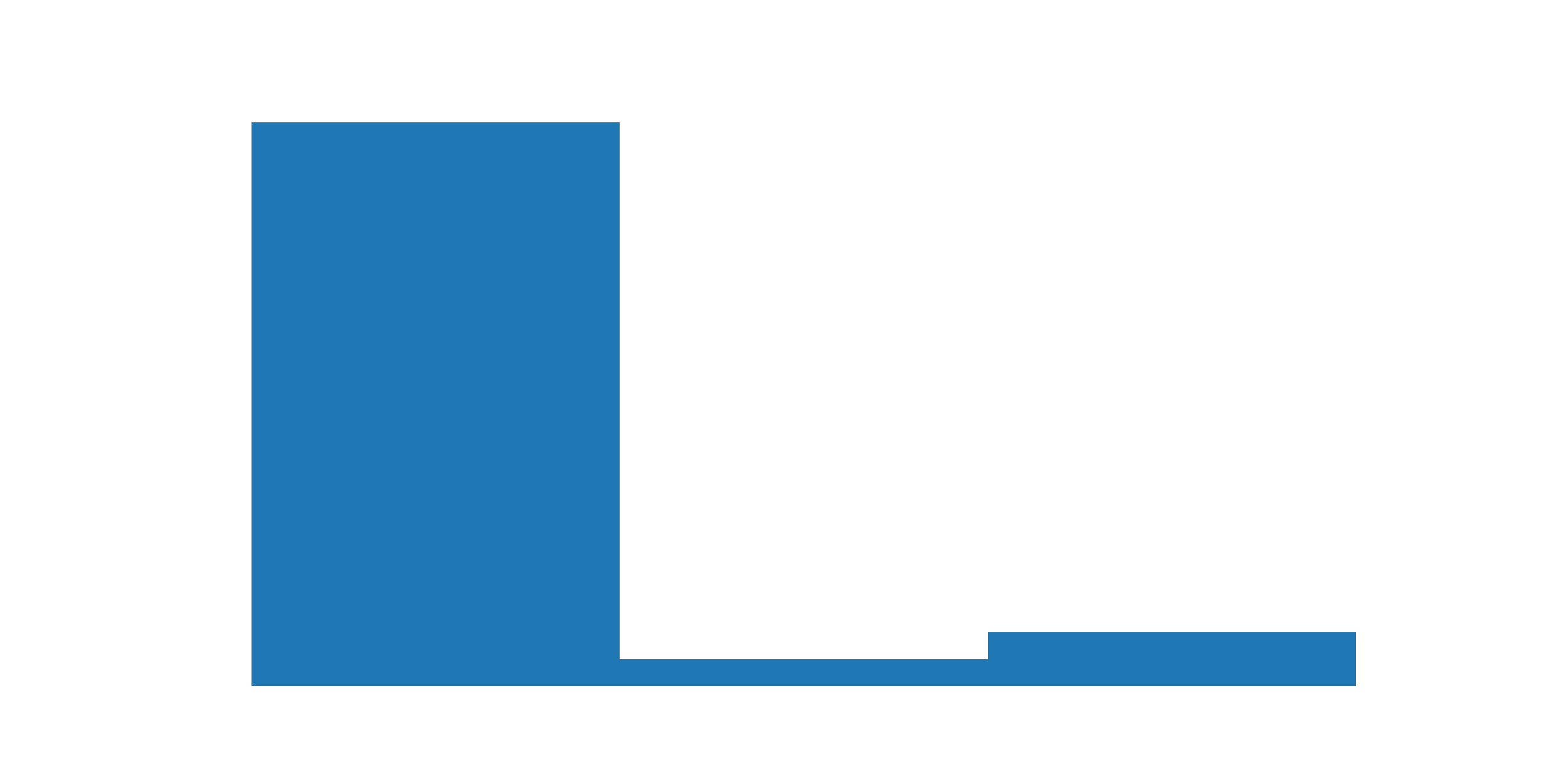}& 0.42 & 0.04 & 0.91 \\
1984&Economic Modelling & 20 & \zerofiveinx{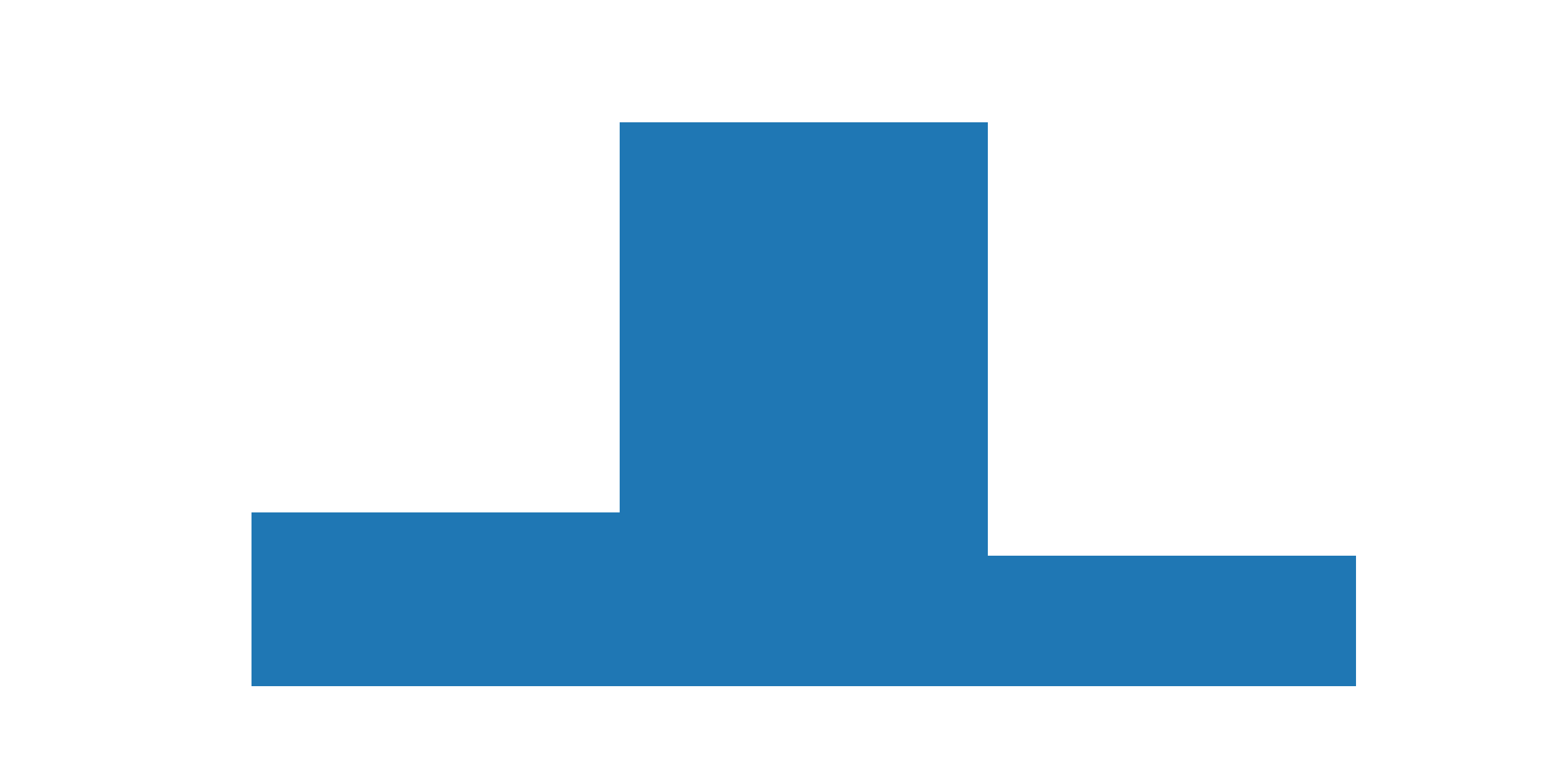}& 0.50 & 0.05 & 0.91\\  
1973&Energy Policy& 75 & \zerofiveinx{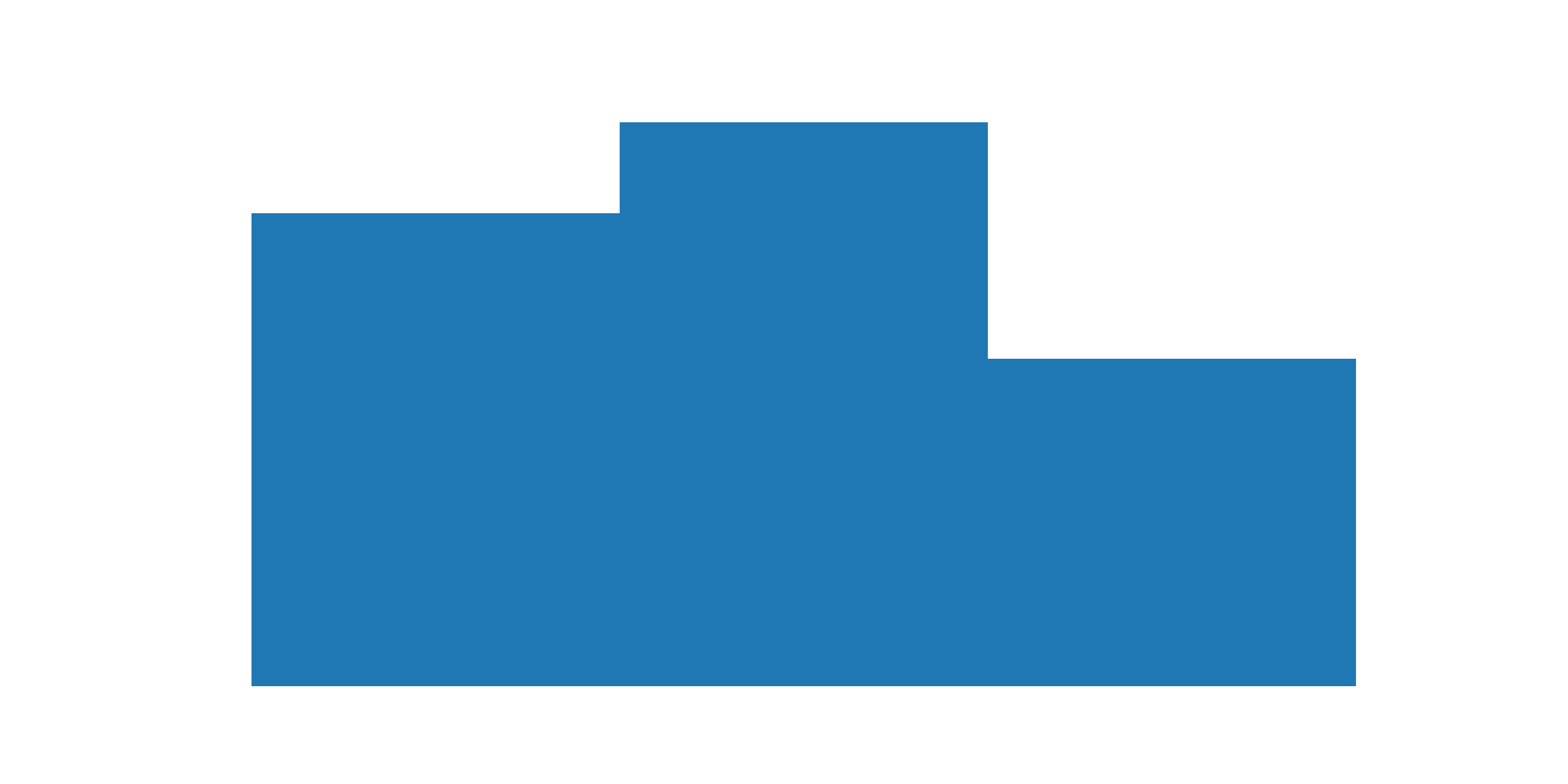}&0.64 & 0.28 & 0.70  \\
1979&Energy Economics& 52 &\zerofiveinx{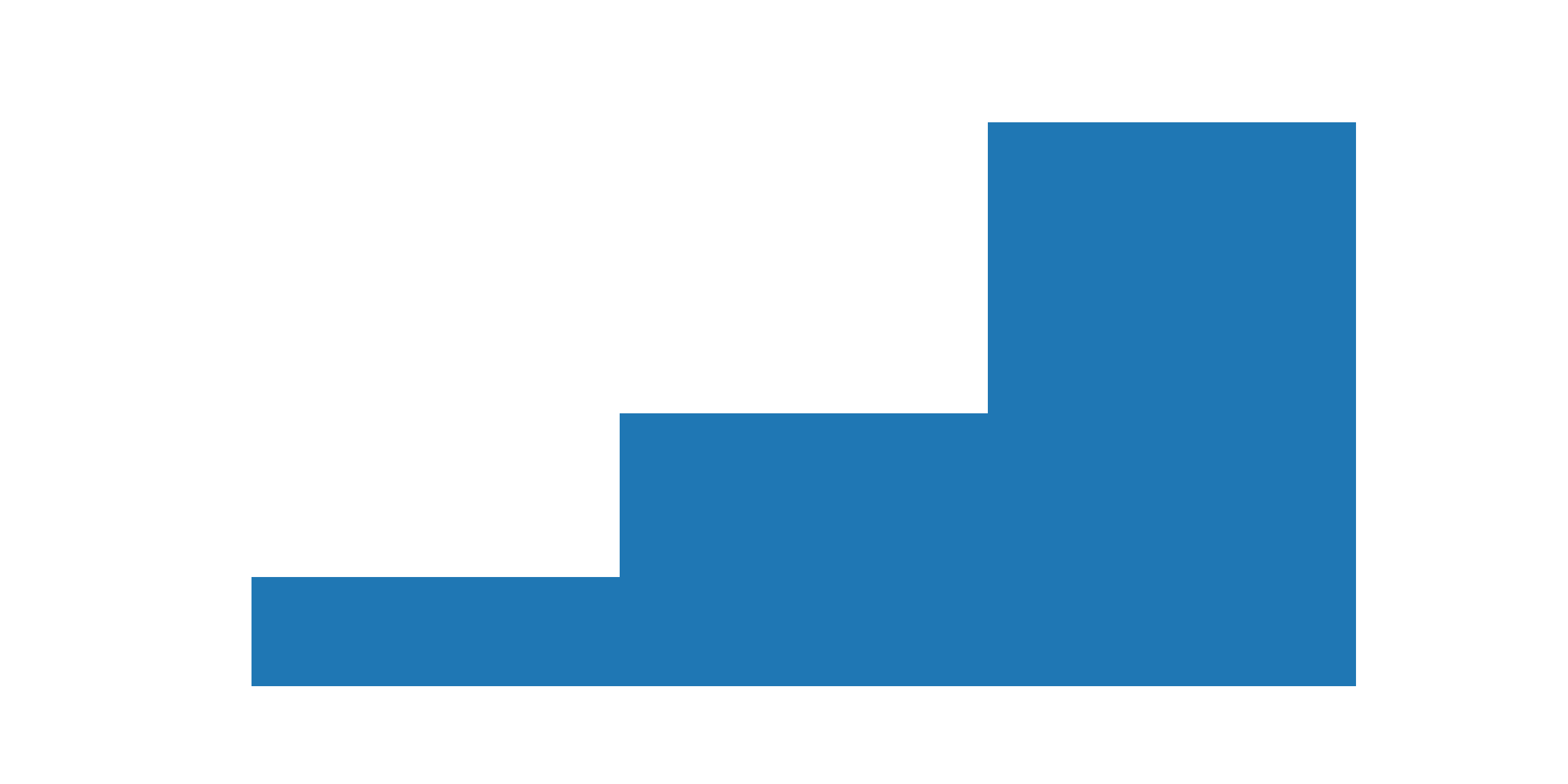}& 0.50 & 0.31 & 0.62   \\
1999&Environ. Dev. Sustainability & 28 & \zerofiveinx{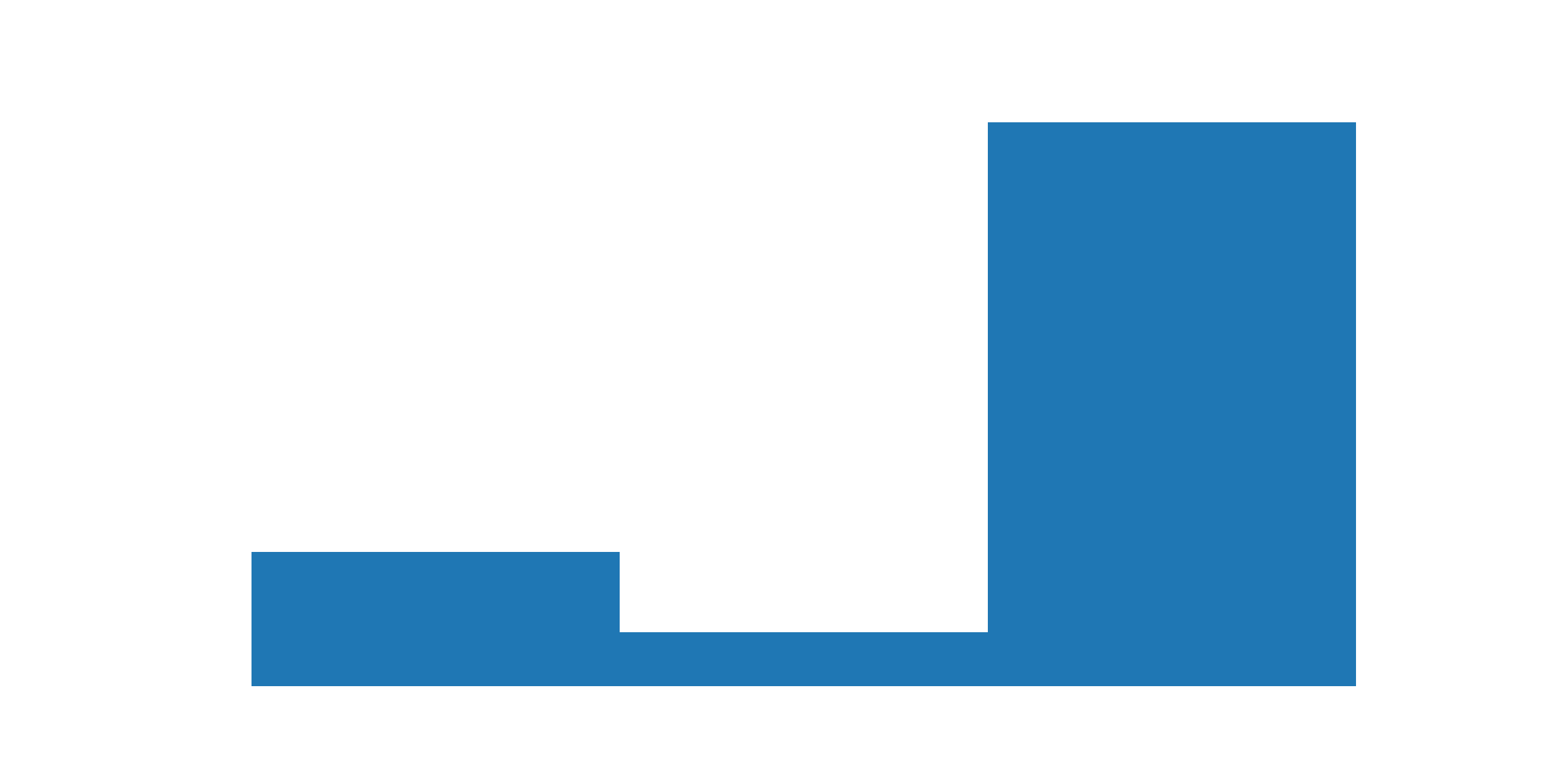}& 0.54 & 0.39 & 0.58  \\
1970&J. Environmental Management & 20 & \zerofiveinx{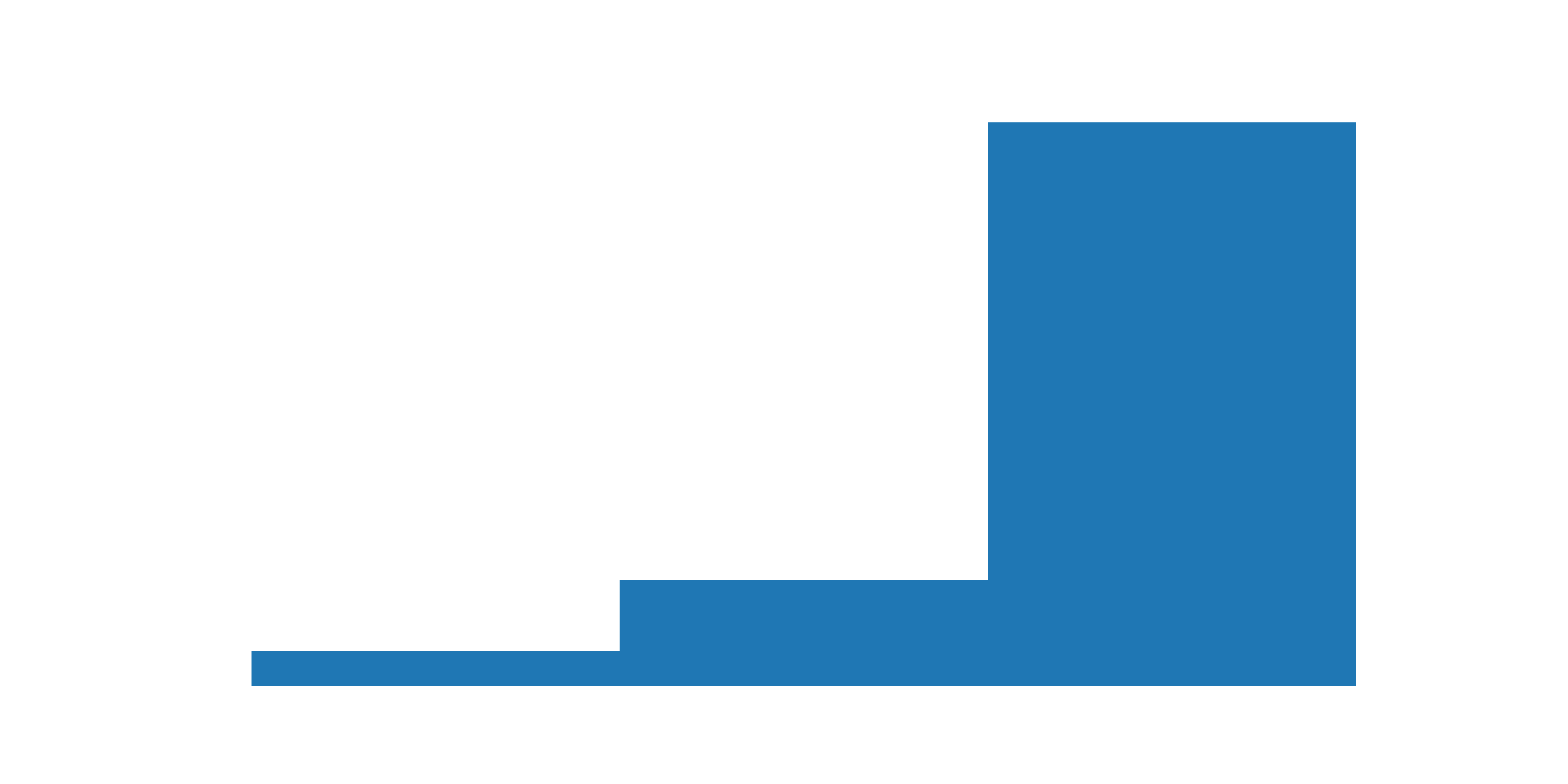} & 0.40 & 0.30 & 0.57 \\
1997&Renew. Sustainable Energy Rev & 61 & \zerofiveinx{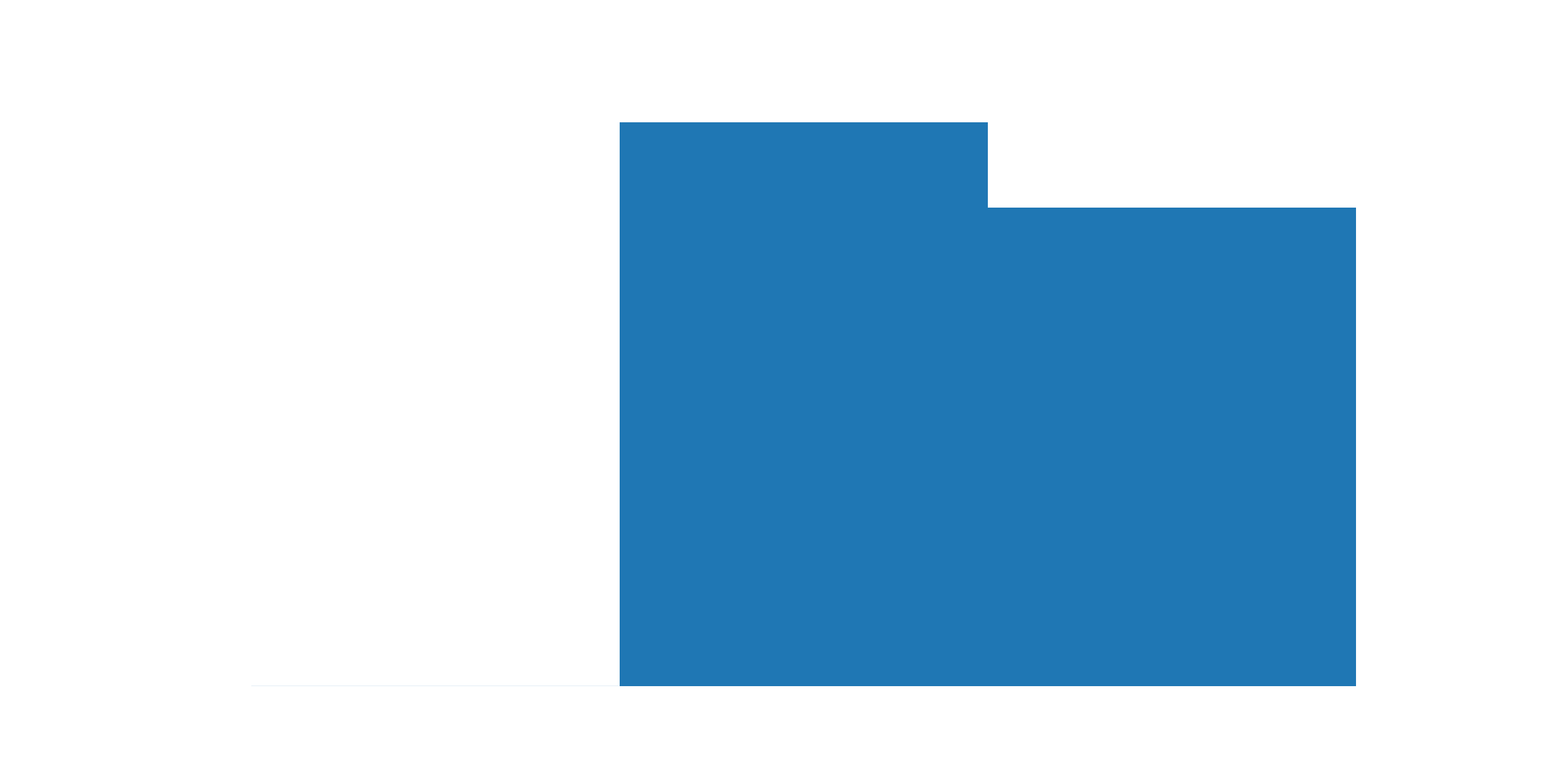} & 0.61 & 0.56 & 0.52 \\
2009&Sustainability & 78 & \zerofiveinx{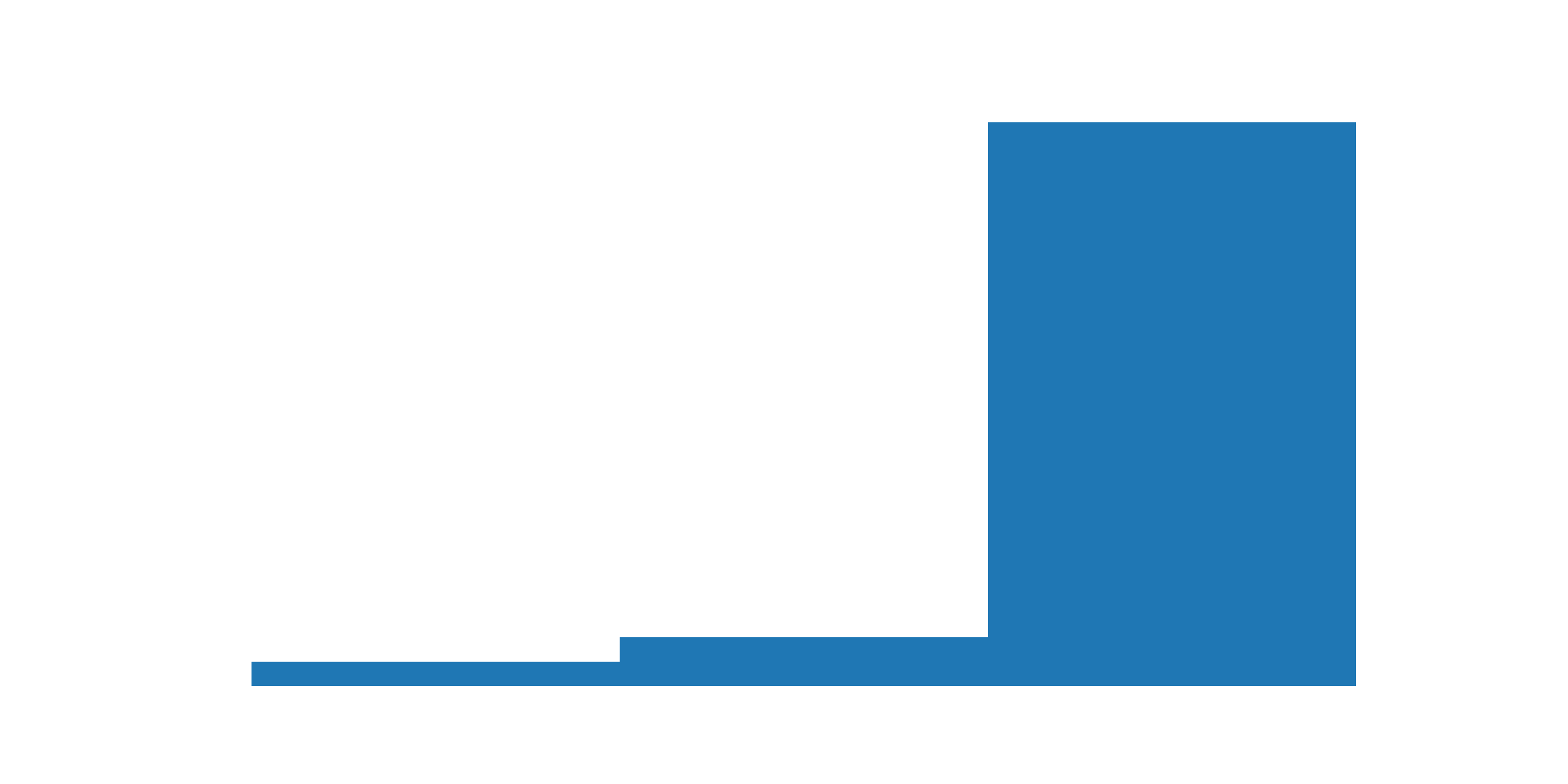} &0.35 & 0.40 & 0.47  \\
1976&Energy & 30 & \zerofiveinx{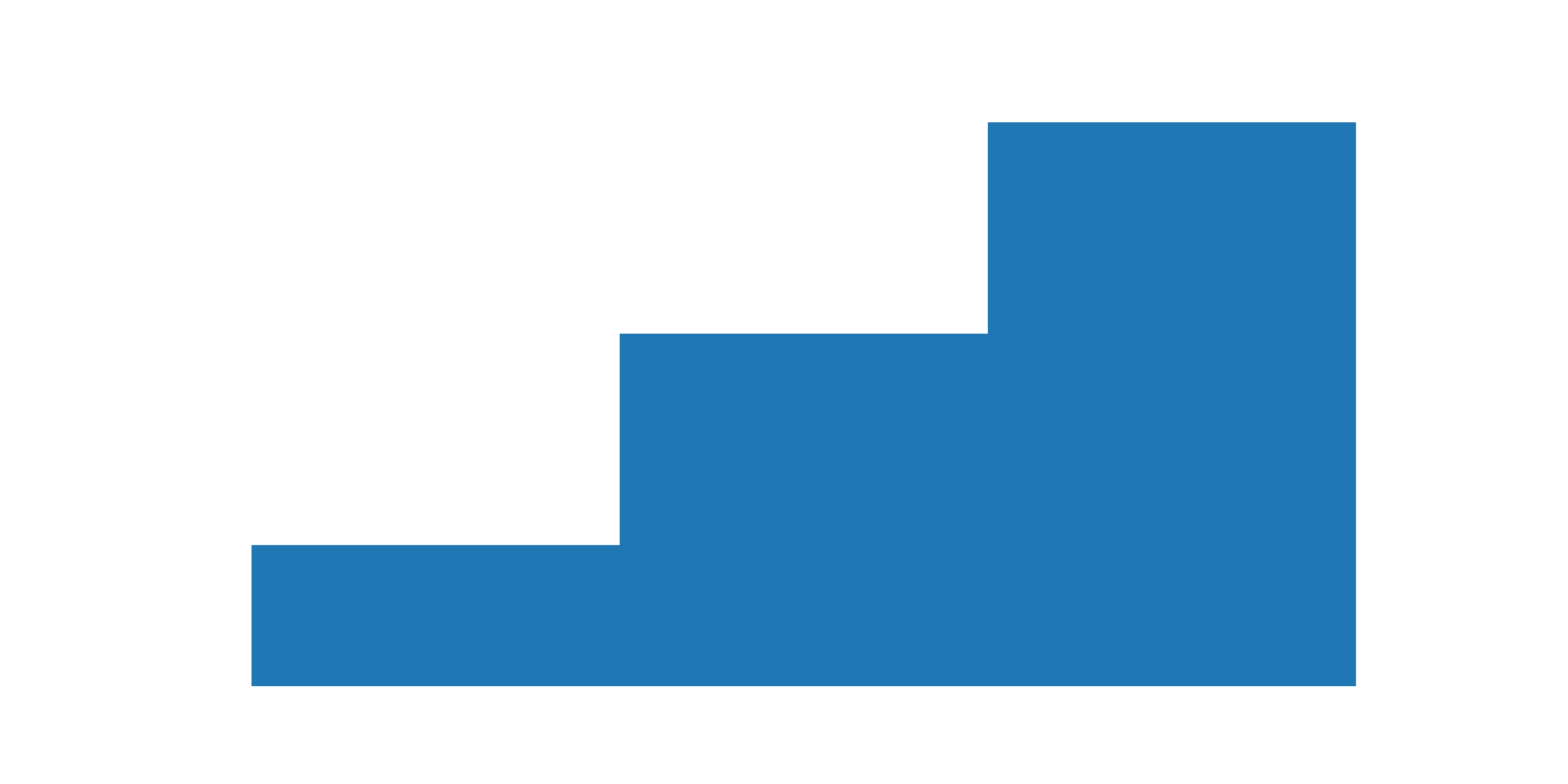} & 0.40 & 0.50 & 0.44 \\
2001&Ecological Indicators & 49 &\zerofiveinx{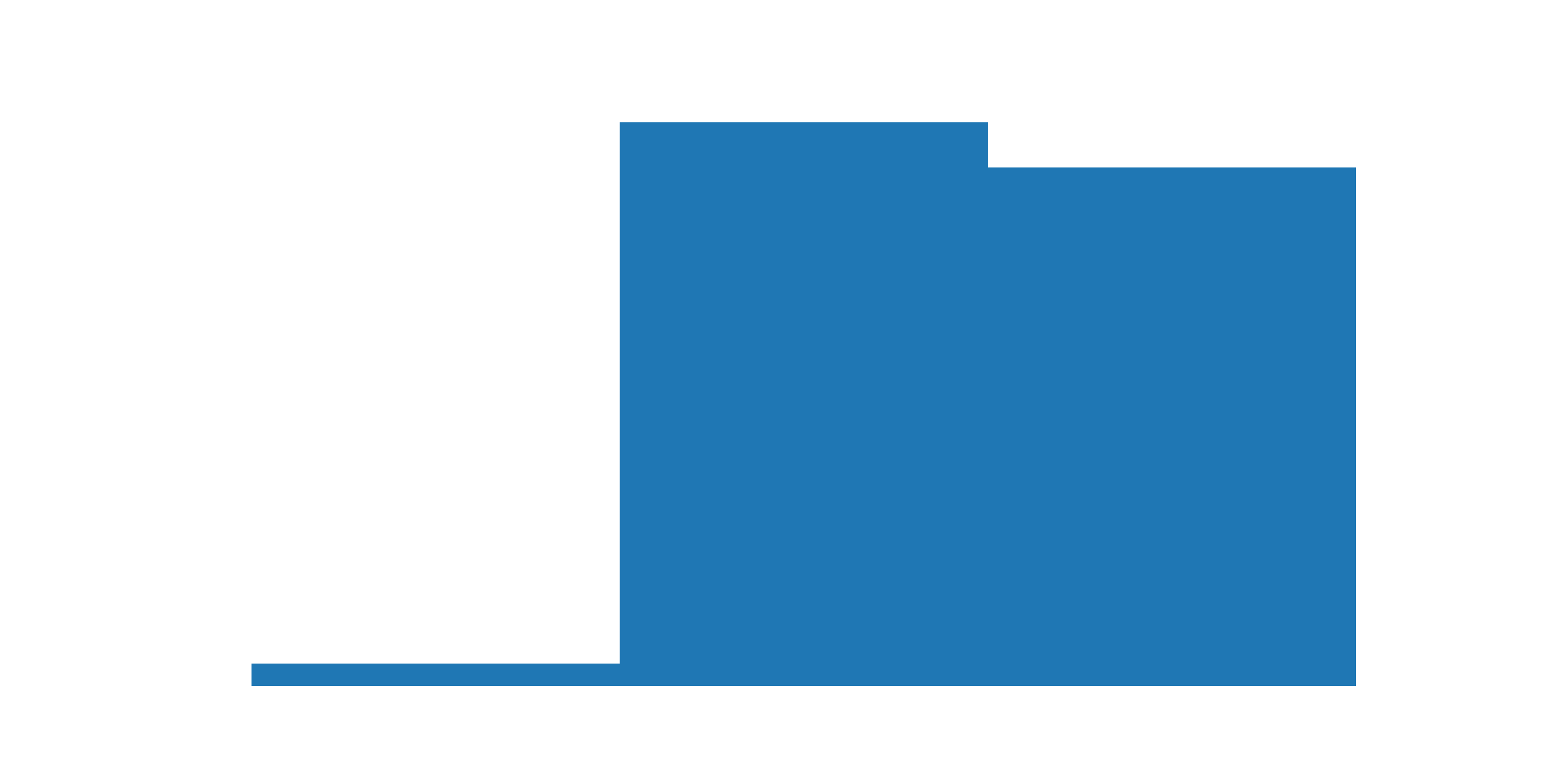}& 0.41 & 0.57 & 0.42  \\
2008&Energies & 23 &\zerofiveinx{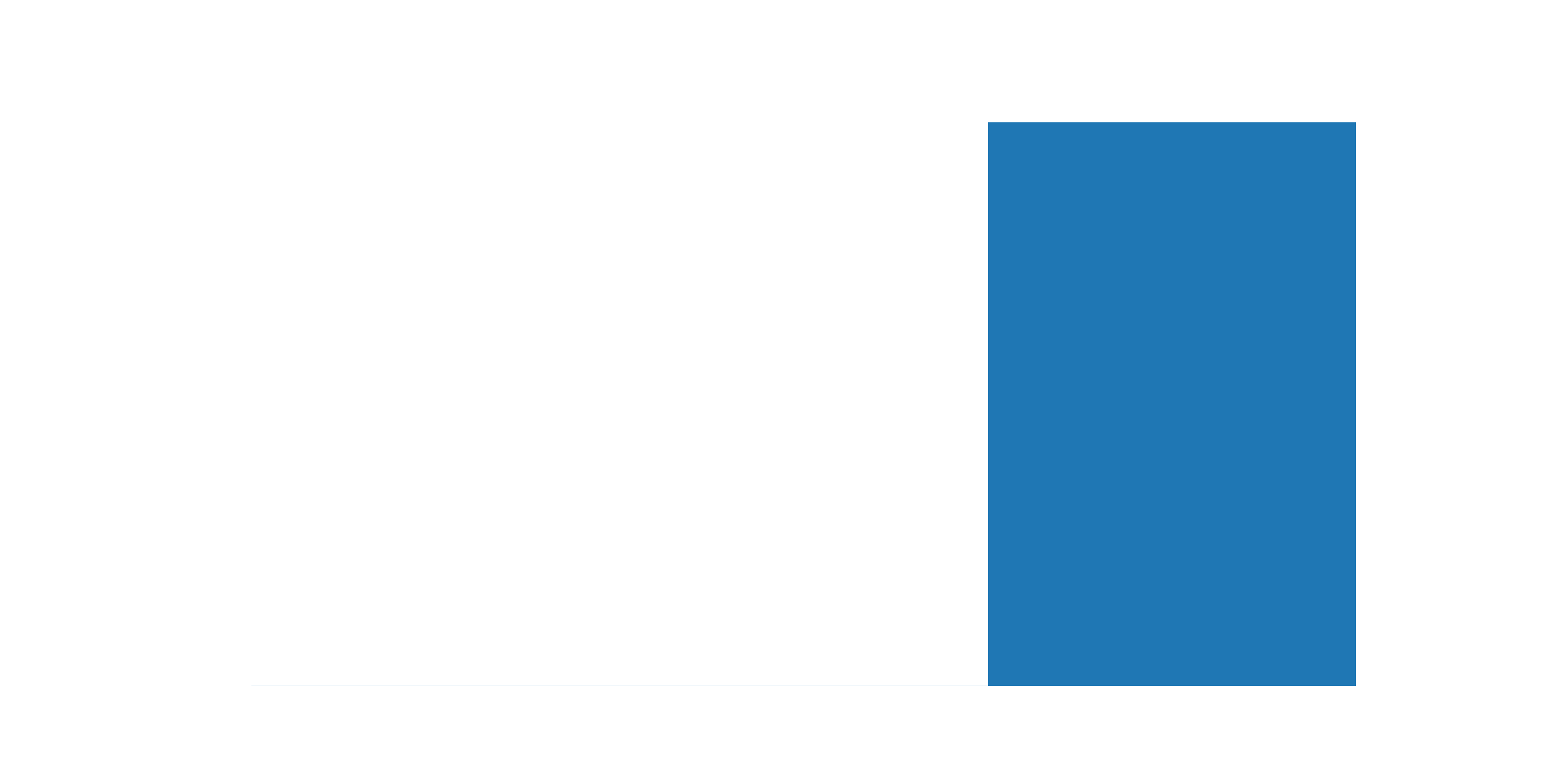} & 0.35 & 0.52 & 0.40 \\
1993&J. Cleaner Production& 102  &\zerofiveinx{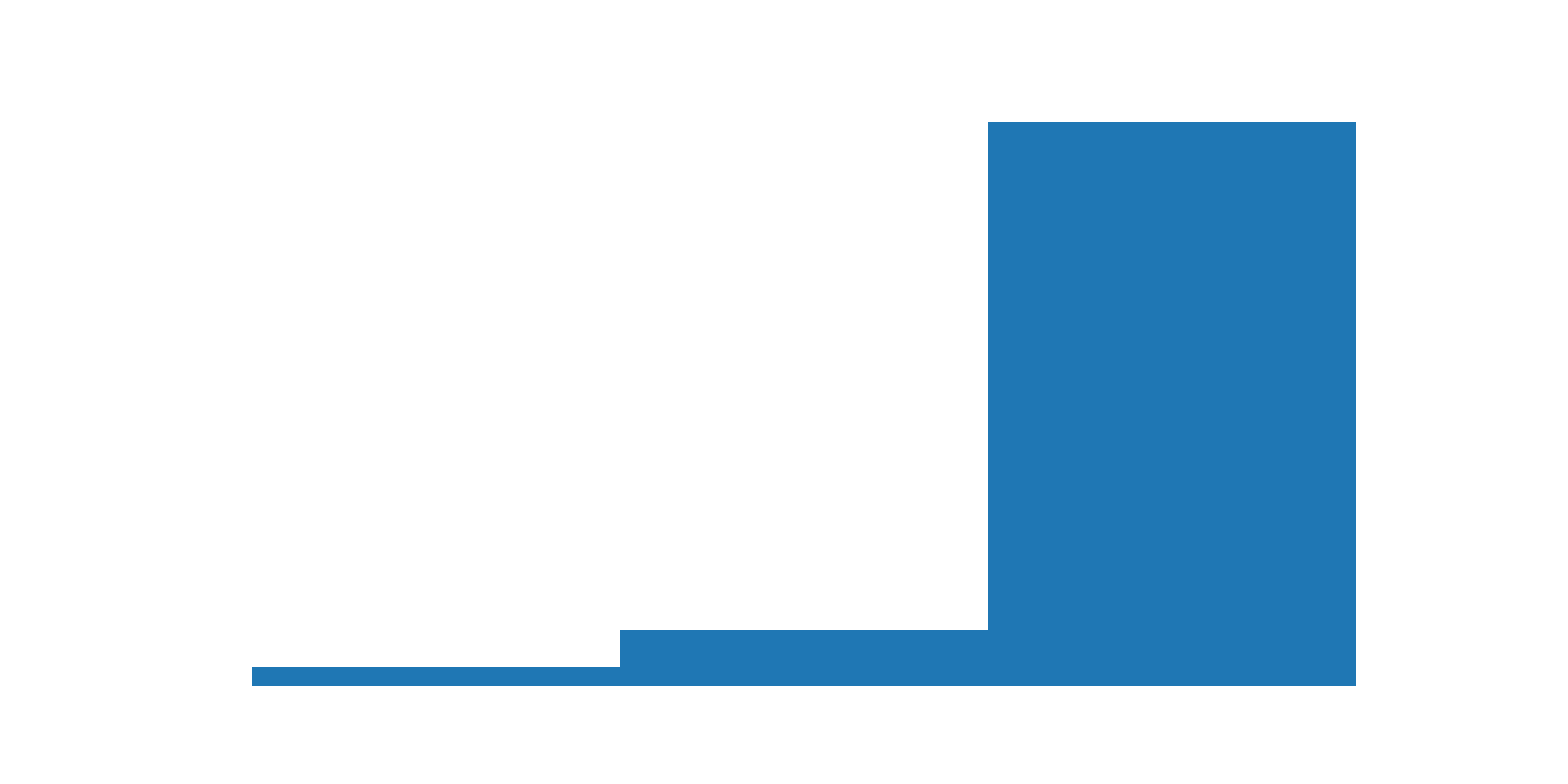} & 0.30 & 0.49 & 0.38 \\
2011&Int. J. Energy Econ. Policy & 57 &\zerofiveinx{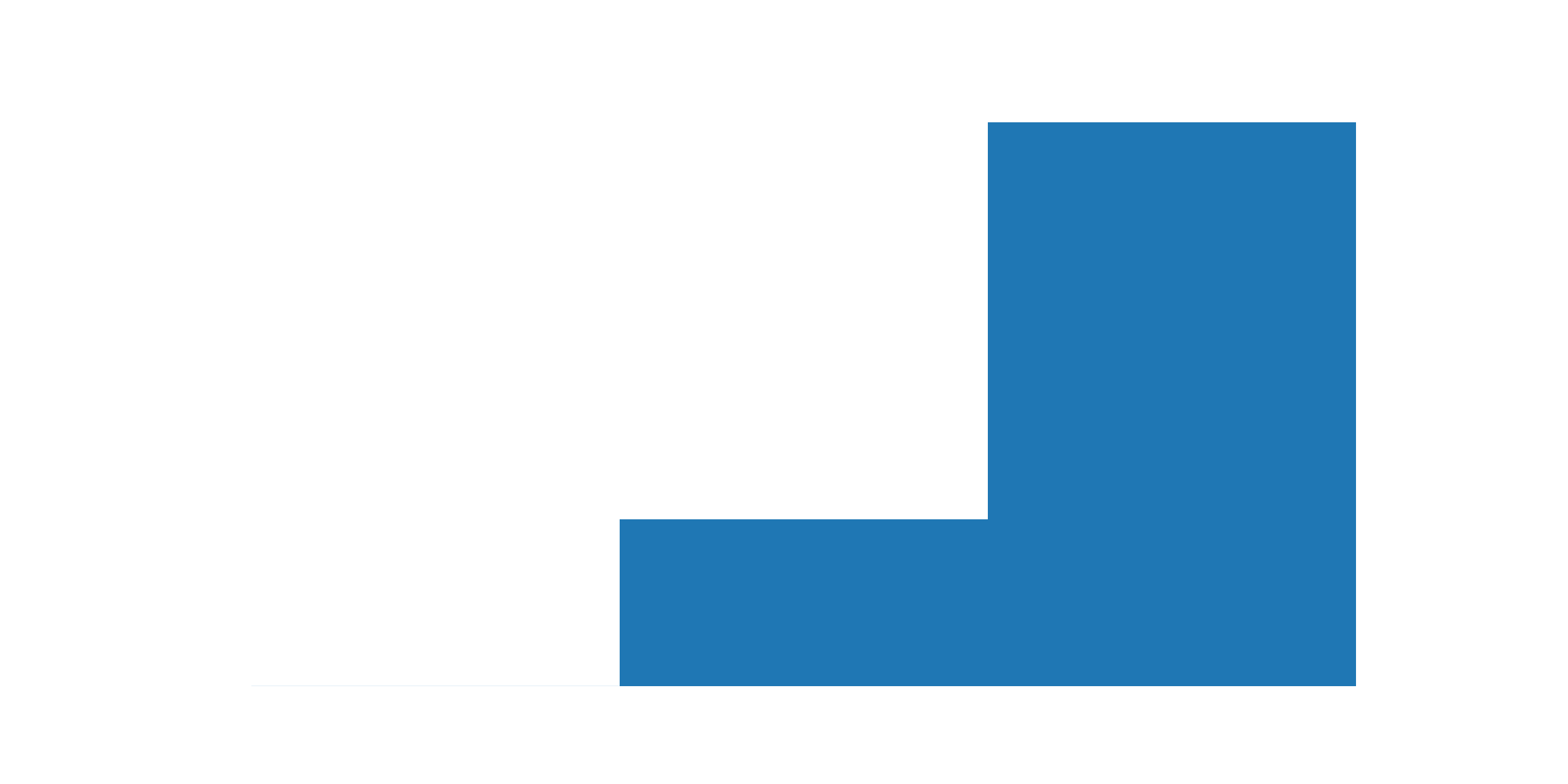}& 0.39 & 0.77 & 0.33 \\
1994&Environ. Sci. Pollution Res. & 286  &   \zerofiveinx{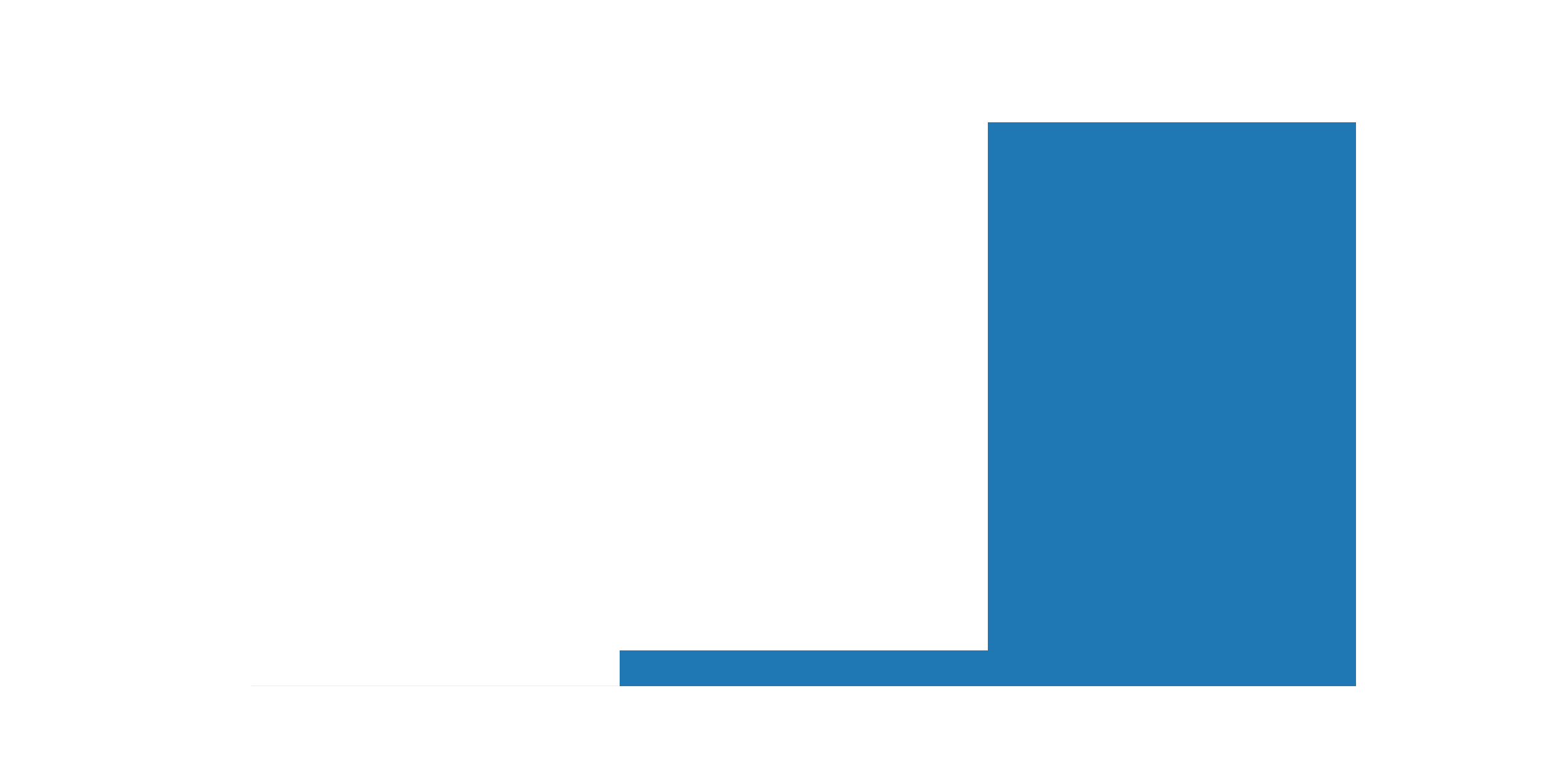}   & 0.30 & 0.72 & 0.30\\
1972&Sci. Total Environment & 52 &\zerofiveinx{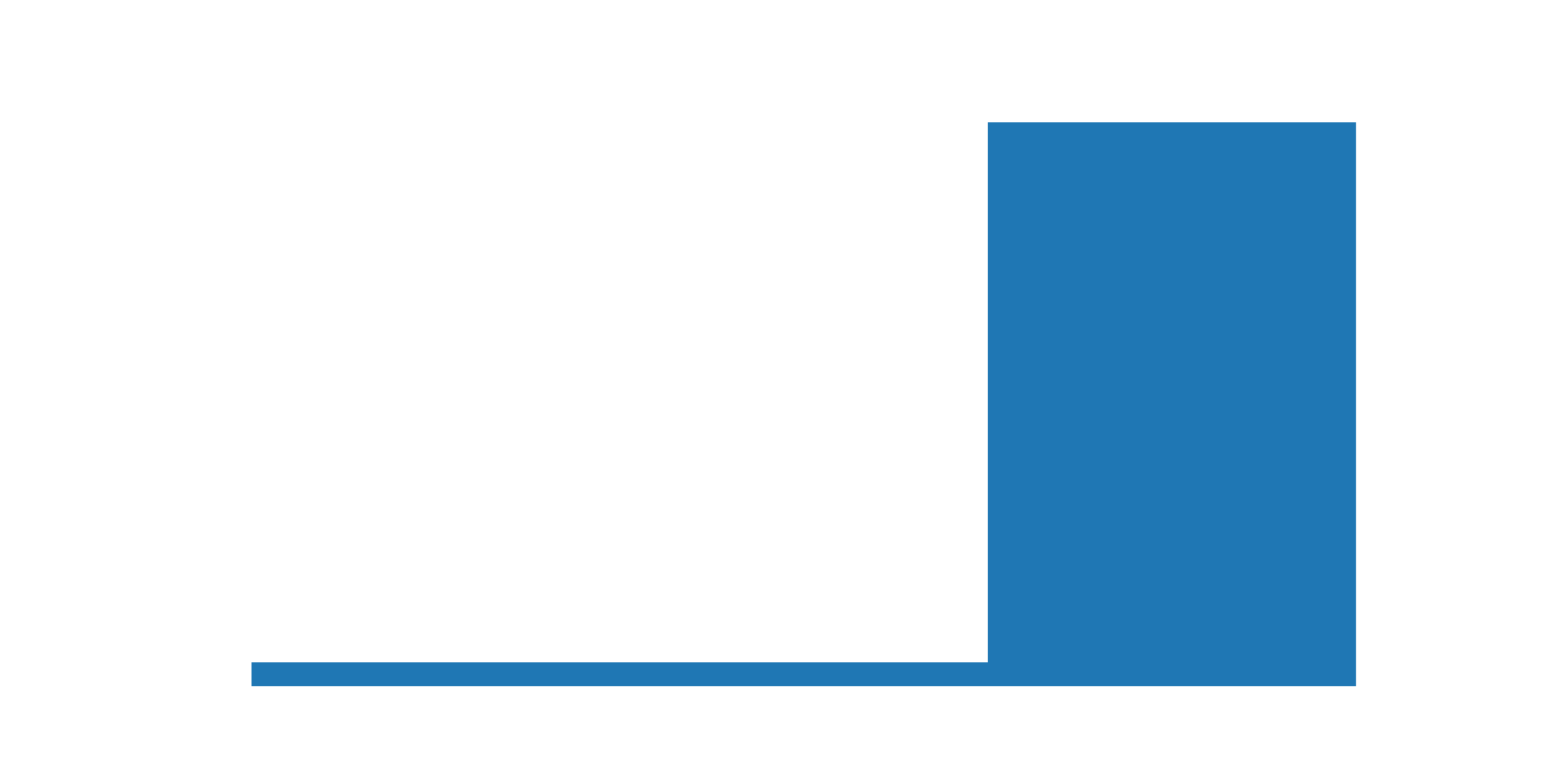} & 0.17 & 0.50 & 0.26\\
\bottomrule
\end{tabularx}
\caption{Journals publishing at least 20 articles, preceded by year of foundation (first volume). \#A indicates the total number of articles in the corpus. The proportion of articles citing respectively Stern and Okturk is shown, as well as the ratio $r$ of Stern to Stern + Öztürk citations, which is used to rank the table. Temporal profiles are bar charts of the number of articles published over the three periods (1995-2011, 2012-16, 2017-21). All profiles are scaled to their maximum value.}
\label{most-publishing-journals}
\end{center}
\end{table}

These temporal profiles deliver an interesting picture. The debate on EKC has originated within studies of development as evidenced by the early importance of a journal such as {\em Environment and Development economics}. It has further taken roots in environmental or energy economics with journals including {\em Environment and Resources Economics}, {\em Ecological Economics} or {\em Energy Policy}. If these journals still publish articles on EKC, the literature from the 2010s is more and more published in journals that are engineering or natural sciences oriented, such as {\em Environmental Science and Pollution Research} or {\em Science of the Total Environment}. 

These temporal profiles signal a shift in journals. The move away from {\em Ecological economics} is especially clear: 80\% of this journal's articles on EKC are published in the first period, before 2011, whereas the field has experienced a tremendous growth afterwards. This is also surprising given the fact that the EKC was identified by a citation analysis of \citet{Ma2006} as a prominent topic of ecological economics (as represented by the journal {\em Ecological economics}), as opposed to environmental economics (as represented by the {\em Journal of Environmental Economics and Management}). In the period 2006-2010, 46 articles of our corpus were indeed published in {\em Ecological economics}, but the flow of publications strongly diminishes afterwards, and the output of EKC research was directed to other journals, and actually outside the sphere of economic research. We can see that move if we classify journals in Table \ref{most-publishing-journals} as economic journals, when the matter classification of the journal in Scopus contains the word ``economics'', and non-economic journals otherwise\footnote{For the journals of Table \ref{most-publishing-journals}, the journals classified as economic are these with 'economic*' in the title, except for {\em Environ. Dev. Sustainability}, which is also classified as economic. }. For the articles published in journals of Table \ref{most-publishing-journals}, 72\% were published in economic journals over the first period 1995-2011, 26\% over the period 2012-2016 and only 11\% over the last period 2017-2021. The move towards non-economic journal is even more pronounced when one zooms in the first period: 96\% of the articles published in journals of Table \ref{most-publishing-journals} were in economic journals over the period 1996-2000, 84\% over the period 2001-2005, and 63\% over the period 2006-2010. The EKC research thus changes significantly over the whole period and was found more and more relevant for fields other than economics. The growth of EKC articles that we have observed is actually attributable to journals classified as non-economic. The field has actually expanded outside economics.

\begin{figure}[!t]
\begin{center}
\includegraphics[width=\linewidth]{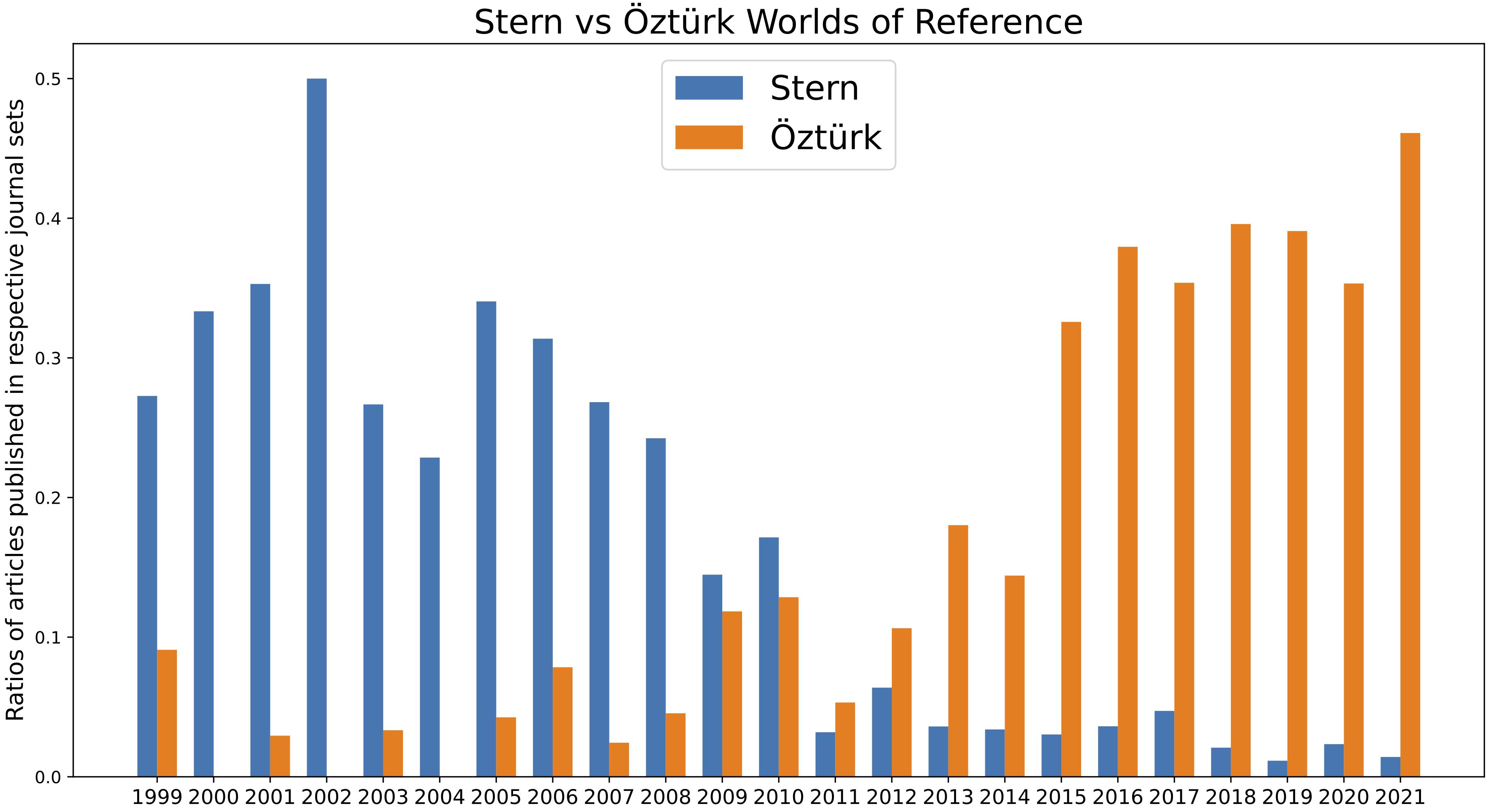}
\caption{Relative number of articles within Stern's and Öztürk's worlds of references. The chart starts in 1999 as there are very few articles before that year and they are all published in Stern's world.}
\label{fig-stern-Öztürk-world-of-ref}
\end{center}
\end{figure}

To get a better sense of this evolution, we go back to Stern and Öztürk. Table~\ref{most-publishing-journals} indicates the proportion of articles citing either Stern and Öztürk for each journal, as well as the ratio $r\in[0,1]$ between Stern and Stern+Öztürk proportions --- a ratio close to 1 indicates a dominance of articles citing Stern. The table is actually ordered by descending value of $r$. Articles in journals on top, mostly environmental economics journals, cite far more Stern than Öztürk, whereas articles on the bottom cite far more Öztürk. The shape of temporal profiles exhibits a strong correlation with $r$ in the sense that higher $r$ ratios visibly correspond to earlier periods of EKC publications.

Furthermore, we define the \emph{world of references} of Stern and of Öztürk: If Stern cites an article that is published in journal J, then J is in the world of references of Stern, and similarly for Öztürk. This aims at grasping the journals that an author is aware of and that he considers worth citing.
Simply examining the sets gives interesting results because for example some journals like \emph{\hbox{J. Cleaner Production}} or \emph{Renewable Sustainable Energy Rev} are absent from Stern’s world of references. The temporal count in figure~\ref{fig-stern-Öztürk-world-of-ref} is even more striking. We compute the yearly distribution of the ratio of articles published in each world of references relative to the total number of articles published that year in our corpus. We can map the sliding of the field as more and more articles are published in Öztürk’s world of references and less and less in Stern’s one. This reflects the growing contribution in the last time period of journals such as \hbox{\em Environ. Sci. Pollution Res.} or \emph{Sustainability} (from the controversial publisher MDPI) that are principally active in the last period yet are also not in Stern’s world of references.
Finally, this imbalance does not concern only the two authors under scrutiny but also extends to journal publication patterns observed in the broader blocks. Table~\ref{tab:wor-blocks} shows the share of articles that are published by authors of each block in the respective Öztürk \hbox{vs.} Stern world of references journals: blocks C and D almost do not publish in Stern journals, and do publish massively in Öztürk journals, whereas block A principally publishes in Stern journals and block B, interestingly, exhibits a balanced diet of publication between Stern and Öztürk journals.
\begin{table}[!t]
\begin{center}
\begin{tabular}{>{\em}ccc}
\toprule
Block&``Stern journals''&``Öztürk journals''\\\midrule
A & .571 & .111 \\
B & .161 & .164 \\
C & .007 & .564 \\
D & .022 & .436 \\
\bottomrule
\end{tabular}
\caption{Ratio of articles published in Öztürk and Stern ``world of references'' journals for each block.}
\label{tab:wor-blocks}
\end{center}
\end{table}

\section{Discussion}
\label{sec-diverging}\label{sec-discussion}

Our results hint at the coexistence of two main epistemic practices with relatively distinct citation and, accordingly, co-authorship networks, also centered around two distinct periods of time, while a shift exists not only in terms of publication venues but also topics, while the relatively balanced share of positive \hbox{vs.} negative results indicates ongoing debate on the empirical validation of the EKC. This further hints at the co-existence of distinct epistemic communities, which we understand here as a community of actors who share a goal of knowledge creation within a common set of topics, beliefs, rules, and, in fine, practices. This notion of epistemic community builds upon two main strands: in political science, where the concept has been introduced \citep{rugg-inte,haas-intr} and in social epistemology \citep{schm:soci,gier:scie}, whereby an epistemic community corresponds to a group of scientists who acknowledge a specific collection of conceptual tools and representations, within a ``paradigm'', may collaborate on specialized tasks, possibly working in a decentralized manner. The joint notion of a shared subset of knowledge issues and procedural authority that characterizes epistemic communities \citep{cowa-expl} is particularly relevant in our context, where we focus on a main issue (the EKC), various subtopics, and diverse blocks of actors defined by shared citation or reference patterns \citep{cohe:orga}.

We would like to offer several discussion points that also relate to some limitations of our approach.
For one, we see here that the algorithm tends to under-report negative results. Despite the correction that we apply, this sheds light on a broader issue.
In effect, articles questioning the rationale behind the EKC and the methods used to find it are not classified as negative results, even though they could be regarded so as they express reservations on the possibility of existence of the object under investigation.

Some further caution is warranted to interpret the number of positive \hbox{vs.} negative results. The following remarks do not address the link between the sentences in the abstract and what is found by the algorithm, but rather the various links between the research and the sentences of abstracts.

First, there is a general tendency in research to over-publish significant results \hbox{vs.} null results \citep[e.g.,][]{Sterling1959,Sterling1995,Ferraro2020}. Null results here would mean the absence of any meaningful relationship between, say, GDP and a pollutant. In the EKC context, however, the publication bias for positive results does not seem relevant: given how the very existence of an EKC for various pollutants has been framed from the very beginning, null results are likely to be conceived as results that invalidate the EKC hypothesis and, thus, published.

Second, there is a reporting bias in the text itself, namely a potential discrepancy between what is found by the authors in the article and what is reported in the abstract. The abstract may report positive results whereas the article actually finds mixed results.  A somehow different example of reporting bias is found in \citet{Apergis2015} (cited 187 times, 9th most cited article). The abstract reports ``empirical support to the presence of an Environmental Kuznets Curve hypothesis''. However, the article investigates the EKC thanks to a regression involving up to the third power of GDP. The coefficient of the first power is positive and coefficient of the second power is negative, hence the claim that ``the estimates have the expected signs'' for an EKC. However, the coefficient of the third power is positive, which shows the presence of an N-shape, usually considered a refutation of an EKC. The article offers no explanations of why the third power has been overlooked.

It is also interesting to contrast the methodology followed and the results it delivers with an assessment from an expert in the field.
Given his activity time span and dominance in citations patterns coming from all blocks, as the analysis has shown, David I. Stern may be considered the leading expert.
As stated in his recent review, \cite{Stern2017} is quite critical of the EKC hypothesis. He shows raw data on emissions of \CO{} per capita and GDP per capita and concludes ``there is little sign of an EKC effect'' \cite[18]{Stern2017}.  Following the argument already developed in \citet{Stern2010}, he estimates that ``the relationship between the levels of [...] \CO{} emissions and income per capita is monotonic when the effect of the passage of time is controlled for'' \cite[15]{Stern2017}. A decreasing trend is thus attributed to a time-effect. This skepticism towards EKC reflects a long-standing position. Indeed, the first article \citep[][which was not concerned by \CO]{Stern1996} already emphasizes the mixed results of the empirical investigation of the EKC hypothesis, as well as the estimation problems. 

In this regard, the longest-lasting leading expert in the field would appear to support a view that goes against numerous positive results from the recent literature. There could be several explanations to explain this discrepancy.
One would be that the expert holds a minority position in the field, but another one would be that there are, in fact, two fields.
As we have seen in the previous section, there have been changes in the recent literature. The increase in articles on EKC in the last decade comes from a different breed of journals, with a topical focus that appears to be somewhat distinct from the earlier wave of journals.

Furthermore, our query on Scopus makes the implicit assumption that the term “environmental Kuznets curve” is well-defined in the literature, so that all articles employing it actually speak of the same object. The term seems to be quite narrow, so that we can have at first sight the impression that the assumption is truly warranted. However, a more detailed analysis shows that this is actually questionable.
At the very beginning, the EKC was envisioned as a bivariate relationship between GDP and environmental pressure (depending on pollutant). This is what is reflected in most of the definitions, as the one we provided, that insist on the relationship between GDP on the one hand and environmental degradation on the other. It is also this framing that has been adopted constantly by Stern \citep{Stern1996,Stern2004}, more recently see the definition in \citet[p.~8]{Stern2017} and the standard EKC regression model \citep[p.~13]{Stern2017}. 

Notwithstanding, there have been considerable changes over the three decades that span the scientific existence of the subject. 
An examination of the articles that found an EKC for \CO{} shows that they rarely consider a bivariate relationship between \CO{} and GDP.  A large number of articles are more generally concerned with what they called the emissions / energy and growth nexus. That is, they considered the multivariate relationship between energy, emissions and GDP at least. For example \citet{Apergis2009} consider the causal relationship between \CO{} emissions, energy consumption, and output. Within this relationship, the relationship between carbon dioxide and GDP, with energy consumption controlled for, is still called an EKC when it is of an inverted U shape. The validity of the EKC is thus tested within a setting different from the one framed by Stern, and the meaning of the EKC has changed.

More generally,  further control variables have been introduced, economic or political (institutions), which singularly complicates the meaning and the interpretation of the EKC. New statistical methods, new datasets, new control variables, new environmental stressors, all have stretched the object EKC in different directions, to the point of strongly diminishing its consistency.

The multiple methods available to assess the existence of the EKC also add to the confusion. For example, for the already discussed study on EKC for \CO{} in Vietnam, \citet{Almulali2015} did not confirm the EKC hypothesis, because the elasticity of \CO{} emission with respect to GDP is positive both the short and long run. They rely on the framework set out by \citet{Narayan2010} who have however a different understanding of what constitutes a validation of the EKC. For them, the EKC is validated as long as the long-run elasticity is lower than the short-run elasticity.
One can easily fall into traps when interpreting this kind of results.

In this regard, the blocks and waves that we exhibited hint at the possibility that what we called the field of the EKC research may actually be subdivided into subfields that share the use of the term ``EKC'' and references to seminal articles, yet may diverge in what is meant by it. Thus, our findings on the position occupied by Stern put his assessment into perspective.

Conversely, Stern's assessment of the field and the justifications he put forward points to some limitations of the methods employed here. When counting positive and negative results, we implicitly accept that the underlying articles have used a sound and meaningful method to reach their results, that is that the positive or negative claims made in the abstract are actually genuine results. David I. Stern has however constantly emphasized econometric problems in estimating the EKC. \citet{Stern2001} find that results of regressions are likely spurious and that mispecifications are pervasive. In 2004, he concluded that ``the EKC results have a very ﬂimsy statistical foundation'' \citep{Stern2004}. According to him, the field is still plagued with ``the naïve econometric approaches used in much of the literature'' \citep[p.~24]{Stern2017}. This view is not unique to David I. Stern and can also be found, for example, in \citet{Carson2010} the unique article on the EKC that was published in the {\em Review of Environmental Economics and Policy}, an official journal of the American and European associations of resource and environmental economics.

This reminds that our semantical analysis detects claims, and that claims and results cannot be equated without a critical evaluation of the methodology that substantiates these claims, an evaluation that can usually only be provided by field experts.
Indeed, the critical assessment by Stern goes beyond the impression we can get as external observers that some results may not be reliable – for instance by employing flawed econometric tools as was pointed by \citet{Wagner2008,Wagner2015}.
An external assessment has to factor in this reliability, albeit this is very difficult to do without being an expert of the techniques employed in the literature. In our perspective, it is outside of the scope of this article to develop systematic rules to appraise the methodological aspects of articles on an automated basis.   

In a way, the evolution that we observed on Figure~\ref{fig:articles-posneg} essentially comes down to a quantitative confirmation that there is still an active debate between positive and negative claims around phenomena that are said, by authors, to be connected to the concept of the environmental Kuznets curve. We leave to further research the possibility of characterizing in more detail the approaches and techniques that are typically in use in each of the research streams that we identified, and the related semantic shifts on ``EKC''.

\section{Conclusion}\label{sec-conclusion}

Our analysis of the field structured around the use of ``environmental Kuznets curve'' combines two very recent computational methods in an integrated fashion: semantic and network analyses. For the former, the use of semantic hypergraphs enables us to go beyond lexicometric patterns and to extract both positive and negative claims about the validity of the EKC hypothesis from abstracts. For the latter, using degree-corrected stochastic block-modeling reveals the structure of the author citation network as a compact meta-graph made of a few blocks and easily-interpretable connections between them (i.e., meta-graph nodes are blocks of authors and meta-graph edges are sets of citations between members of blocks).

The combination of topological and semantic features, and a variety of other metrics, both temporal and structural, converges on a characterization of the field that reveals, in essence, the existence of two epistemic communities: one roughly centered around Stern, a long-lasting expert of the field, and one around Öztürk, a more recent expert that also currently dominates the field in terms of citation counts. There also appears to be a remarkable temporal and, to a lesser extent, topical discontinuity between these communities. The first wave and epistemic community, centered around the Stern block, is on the whole less positive on EKC, publishes less often and is less endogamic and is more focused on SOx and NOx. The second wave and epistemic community, consisting of the Öztürk block, publishes more positive results and more results overall, is more focused on GHG and energy and is more endogamic. There is also a smaller epistemic community dominated by Chinese authors and focused on China, that appears to be a middle ground between the two waves according to the various metrics. The divergence of the two communities is also apparent in publication venues --- what we call ``worlds of references'' that are quite distinct and whose activity closely follows the two temporal waves. 

We can risk an explanation of the drift between two communities and the shift in main journals we have observed. The research on the EKC that originates in the development and environmental economics community was very intense in the first years, with lively debates about the existence of the EKC and its policy implications. However, after more than fifteen years of research, part of the economic community became suspicious about the EKC, as evidenced for example by the article published by \citet{Carson2010} in the official journal of the American and European associations of resource and environmental economics, that precisely aims at synthetizing lessons for policy-making learned from research of this community. In those circles, EKC research was no longer fashionable as it had been. Newcomers that were attracted by this research, with a different focus on energy and GHG emissions, appear to have somewhat migrated to more engineering-oriented journals, at the crossroads of environmental sciences and energy research. This is the place where the literature currently experiences a massive increase. Whether this increase will continue at the outskirts of economics or whether it foretells a triumphal comeback of the EKC into economics is an open question. The case of the EKC research tells us that research topics have a life of their own, across disciplinary boundaries, that they can survive and even thrive, even when the passions that animated the original players of the field have faded away for scientific reasons.

If we now focus on claims about the EKC, we observe on the whole that the share of positive claims as reported in abstracts has consistently increased over the years, yet remains of the same order of magnitude as negative claims --- the debate is not closed. Furthermore, our appraisal of positive \hbox{vs.} negative claims is based on authors' reporting in abstracts: casual examination of article contents reveals that some positive abstracts correspond to more nuanced results in the article itself, whereby negative findings are intertwined with positive ones, as well as nuances of what counts as an EKC and which variables should be considered (for instance in terms of economic development, pollutants, and sets of countries). Beyond distinct publication and citation practices, this might more broadly suggest that the discontinuity and difference in results may be related to different understandings of the EKC, and the scientific areas, methodologies and topics that are relevant to its appraisal and validation.

\paragraph{Acknowledgments.}
This research has been partially supported by the “Socsemics” Consolidator grant funded by the European Research Council (ERC) under the European Union Horizon 2020 research and innovation program, grant agreement No. 772743.

\bibliographystyle{OEconomia_EN_2}
\bibliography{ekcbib}

\begin{thebibliography}{67}
\providecommand{\natexlab}[1]{#1}
\providecommand{\url}[1]{\texttt{#1}}
\providecommand{\urlprefix}{URL }
\providecommand{\selectlanguage}[1]{\relax}
\providecommand{\bibAnnoteFile}[1]{%
  \IfFileExists{#1}{\begin{quotation}\noindent\textsc{Key:} #1\\
  \textsc{Annotation:}\ \input{#1}\end{quotation}}{}}
\providecommand{\bibAnnote}[2]{%
  \begin{quotation}\noindent\textsc{Key:} #1\\
  \textsc{Annotation:}\ #2\end{quotation}}

\bibitem[{Acaravci and Ozturk(2010)}]{Acaravci2010}
Acaravci, Ali and Ilhan Ozturk. 2010.
\newblock On the relationship between energy consumption, {CO2} emissions and economic growth in {Europe}.
\newblock \emph{Energy}, \emph{35}(12): 5412--5420.
\bibAnnoteFile{Acaravci2010}

\bibitem[{Ahmed et~al.(2011)Ahmed, Neville, and Kompella}]{ahmed-2011-network}
Ahmed, Nesreen, Jennifer Neville, and Ramana~Rao Kompella. 2011.
\newblock Network Sampling via Edge-based Node Selection with Graph Induction.
\newblock Purdue University.
\bibAnnoteFile{ahmed-2011-network}

\bibitem[{Al-Mulali et~al.(2015)Al-Mulali, Saboori, and Ozturk}]{Almulali2015}
Al-Mulali, Usama, Behnaz Saboori, and Ilhan Ozturk. 2015.
\newblock Investigating the environmental {Kuznets} curve hypothesis in {Vietnam}.
\newblock \emph{Energy Policy}, \emph{76}: 123--131.
\bibAnnoteFile{Almulali2015}

\bibitem[{Andreoni and Levinson(2001)}]{Andreoni2001}
Andreoni, James and Arik Levinson. 2001.
\newblock The simple analytics of the environmental {Kuznets} curve.
\newblock \emph{Journal of Public Economics}, \emph{80}(2): 269--286.
\bibAnnoteFile{Andreoni2001}

\bibitem[{Ang(2007)}]{Ang2007}
Ang, James~B. 2007.
\newblock {CO2} emissions, energy consumption, and output in {France}.
\newblock \emph{Energy Policy}, \emph{35}(10): 4772--4778.
\bibAnnoteFile{Ang2007}

\bibitem[{Anwar et~al.(2022)Anwar, Zhang, Asmi, Hussain, Plantinga, Zafar, and Sinha}]{Anwar2022}
Anwar, Muhammad~Azfar, Qingyu Zhang, Fahad Asmi, Nazim Hussain, Auke Plantinga, Muhammad~Wasif Zafar, and Avik Sinha. 2022.
\newblock Global perspectives on environmental kuznets curve: {A} bibliometric review.
\newblock \emph{Gondwana Research}, \emph{103}: 135--145.
\bibAnnoteFile{Anwar2022}

\bibitem[{Apergis and Ozturk(2015)}]{Apergis2015}
Apergis, Nicholas and Ilhan Ozturk. 2015.
\newblock Testing environmental Kuznets curve hypothesis in Asian countries.
\newblock \emph{Ecological indicators}, \emph{52}: 16--22.
\bibAnnoteFile{Apergis2015}

\bibitem[{Apergis and Payne(2009)}]{Apergis2009}
Apergis, Nicholas and James~E. Payne. 2009.
\newblock {CO2} emissions, energy usage, and output in {Central} {America}.
\newblock \emph{Energy Policy}, \emph{37}(8): 3282--3286.
\bibAnnoteFile{Apergis2009}

\bibitem[{Barbier(1997)}]{Barbier1997}
Barbier, Edward~B. 1997.
\newblock Introduction to the environmental Kuznets curve special issue.
\newblock \emph{Environment and Development Economics}, \emph{2}(4): 369--81.
\bibAnnoteFile{Barbier1997}

\bibitem[{Bashir et~al.(2021)Bashir, Ma, Bashir, and Shahzad}]{Bashir2021}
Bashir, Muhammad~Farhan, Benjiang Ma, Muhammad~Adnan Bashir, and Luqman Shahzad. 2021.
\newblock Scientific data-driven evaluation of academic publications on environmental Kuznets curve.
\newblock \emph{Environmental Science and Pollution Research}, \emph{28}: 16982--16999.
\bibAnnoteFile{Bashir2021}

\bibitem[{Beckerman(1992)}]{Beckerman1992}
Beckerman, Wilfred. 1992.
\newblock Economic growth and the environment: {Whose} growth? whose environment?
\newblock \emph{World Development}, \emph{20}(4): 481--496.
\bibAnnoteFile{Beckerman1992}

\bibitem[{Bonaccorsi et~al.(2021)Bonaccorsi, Melluso, Chiarello, and Fantoni}]{bonaccorsi2021credibility}
Bonaccorsi, Andrea, Nicola Melluso, Filippo Chiarello, and Gualtiero Fantoni. 2021.
\newblock The credibility of research impact statements: A new analysis of REF with Semantic Hypergraphs.
\newblock \emph{Science and Public Policy}, \emph{48}(2): 212--225.
\bibAnnoteFile{bonaccorsi2021credibility}

\bibitem[{Brock and Taylor(2010)}]{Brock2010}
Brock, William~A. and M.~Scott Taylor. 2010.
\newblock The {Green} {Solow} model.
\newblock \emph{Journal of Economic Growth}, \emph{15}(2): 127--153.
\bibAnnoteFile{Brock2010}

\bibitem[{Brundtland et~al.(1987)Brundtland, Khalid, Agnelli, Al-Athel, Chidzero, Fadika, Hauff, Lang, Shijun, de~Botero et~al.}]{Brundtland1987}
Brundtland, Gro~Harlem, Mansour Khalid, Susanna Agnelli, Sali Al-Athel, Bernard Chidzero, Lamina Fadika, Volker Hauff, Istvan Lang, Ma~Shijun, Margarita~Morino de~Botero et~al. 1987.
\newblock \emph{Our common future}.
\newblock Oxford University Press.
\bibAnnoteFile{Brundtland1987}

\bibitem[{Carson(2010)}]{Carson2010}
Carson, Richard~T. 2010.
\newblock The {Environmental} {Kuznets} {Curve}: {Seeking} {Empirical} {Regularity} and {Theoretical} {Structure}.
\newblock \emph{Review of Environmental Economics and Policy}, \emph{4}(1): 3--23.
\bibAnnoteFile{Carson2010}

\bibitem[{Claveau and Gingras(2016)}]{Claveau2016}
Claveau, Fran\c{c}ois and Yves Gingras. 2016.
\newblock Macrodynamics of {Economics}: {A} {Bibliometric} {History}.
\newblock \emph{History of Political Economy}, \emph{48}(4): 551--592.
\bibAnnoteFile{Claveau2016}

\bibitem[{Claveau and Herfeld(2018)}]{Claveau2018}
Claveau, Fran\c{c}ois and Catherine Herfeld. 2018.
\newblock Network {Analysis} in the {History} of {Economics}.
\newblock \emph{History of Political Economy}, \emph{50}(3): 597--603.
\bibAnnoteFile{Claveau2018}

\bibitem[{Cohendet et~al.(2001)Cohendet, Cr\'eplet, and Dupouet}]{cohe:orga}
Cohendet, Patrick, Fr\'ed\'eric Cr\'eplet, and Olivier Dupouet. 2001.
\newblock \emph{Economics with Heterogeneous Interacting Agents}, Berlin: Springer, chapter Organisational innovation, communities of practice and epistemic communities: the case of Linux.
\newblock 303--326.
\bibAnnoteFile{cohe:orga}

\bibitem[{Cowan et~al.(2000)Cowan, David, and Foray}]{cowa-expl}
Cowan, Robin, Paul~A. David, and Dominique Foray. 2000.
\newblock The Explicit Economics of Knowledge Codification and Tacitness.
\newblock \emph{Industrial \& Corporate Change}, \emph{9}(2): 212--253.
\bibAnnoteFile{cowa-expl}

\bibitem[{Cropper and Griffiths(1994)}]{Cropper1994}
Cropper, Maureen and Charles Griffiths. 1994.
\newblock The interaction of population growth and environmental quality.
\newblock \emph{The American Economic Review}, \emph{84}(2): 250--254.
\bibAnnoteFile{Cropper1994}

\bibitem[{Dinda(2004)}]{Dinda2004}
Dinda, Soumyananda. 2004.
\newblock Environmental {Kuznets} {Curve} {Hypothesis}: {A} {Survey}.
\newblock \emph{Ecological Economics}, \emph{49}(4): 431--455.
\bibAnnoteFile{Dinda2004}

\bibitem[{Ettinger(2020)}]{ettinger2020bert}
Ettinger, Allyson. 2020.
\newblock What BERT is not: Lessons from a new suite of psycholinguistic diagnostics for language models.
\newblock \emph{Transactions of the Association for Computational Linguistics}, \emph{8}: 34--48.
\bibAnnoteFile{ettinger2020bert}

\bibitem[{Ferraro and Shukla(2020)}]{Ferraro2020}
Ferraro, Paul~J. and Pallavi Shukla. 2020.
\newblock Is a {Replicability} {Crisis} on the {Horizon} for {Environmental} and {Resource} {Economics}?
\newblock \emph{Review of Environmental Economics and Policy}, \emph{14}(2): 339--351.
\bibAnnoteFile{Ferraro2020}

\bibitem[{Giere(2002)}]{gier:scie}
Giere, R. 2002.
\newblock Scientific Cognition as Distributed Cognition.
\newblock In Peter Carruthers, Stephen Stitch, and Michael Siegal (eds.), \emph{The Cognitive Basis of Science}, Cambridge University Press. 285--299.
\bibAnnoteFile{gier:scie}

\bibitem[{Grossman and Krueger(1991)}]{Grossman1991}
Grossman, Gene~M. and Alan~B. Krueger. 1991.
\newblock Environmental Impacts of a North American Free Trade Agreement.
\newblock Cambridge (Ma.): NBER.
\newblock Working Paper 3914.
\bibAnnoteFile{Grossman1991}

\bibitem[{Haas(1992)}]{haas-intr}
Haas, P. 1992.
\newblock Introduction: epistemic communities and international policy coordination.
\newblock \emph{International Organization}, \emph{46}(1): 1--35.
\bibAnnoteFile{haas-intr}

\bibitem[{Haberl et~al.(2020)Haberl, Wiedenhofer, Virág, Kalt, Plank, Brockway, Fishman, Hausknost, Krausmann, Leon-Gruchalski, Mayer, Pichler, Schaffartzik, Sousa, Streeck, and Creutzig}]{Haberl2020}
Haberl, Helmut, Dominik Wiedenhofer, Doris Virág, Gerald Kalt, Barbara Plank, Paul Brockway, Tomer Fishman, Daniel Hausknost, Fridolin Krausmann, Bartholom\"aus Leon-Gruchalski, Andreas Mayer, Melanie Pichler, Anke Schaffartzik, T\^ania Sousa, Jan Streeck, and Felix Creutzig. 2020.
\newblock A systematic review of the evidence on decoupling of {GDP}, resource use and {GHG} emissions, part {II}: synthesizing the insights.
\newblock \emph{Environmental Research Letters}, \emph{15}(6): 065003.
\bibAnnoteFile{Haberl2020}

\bibitem[{Herfeld and Malte(2018)}]{Herfeld2018}
Herfeld, Catherine and Doehne Malte. 2018.
\newblock Five reasons for the use of network analysis in the history of economics.
\newblock \emph{Journal of Economic Methodology}, \emph{25}(4): 311--328.
\bibAnnoteFile{Herfeld2018}

\bibitem[{Holland et~al.(1983)Holland, Laskey, and Leinhardt}]{holland1983stochastic}
Holland, Paul~W, Kathryn~Blackmond Laskey, and Samuel Leinhardt. 1983.
\newblock Stochastic blockmodels: First steps.
\newblock \emph{Social networks}, \emph{5}(2): 109--137.
\bibAnnoteFile{holland1983stochastic}

\bibitem[{Jalali et~al.(2016)Jalali, Rezvanian, and Meybodi}]{jalali2016social}
Jalali, Zeinab~S, Alireza Rezvanian, and Mohammad~Reza Meybodi. 2016.
\newblock Social network sampling using spanning trees.
\newblock \emph{International Journal of Modern Physics C}, \emph{27}(05): 1650052.
\bibAnnoteFile{jalali2016social}

\bibitem[{Jia et~al.(2008)Jia, Hoberock, Garland, and Hart}]{jia2008visualization}
Jia, Yuntao, Jared Hoberock, Michael Garland, and John Hart. 2008.
\newblock On the visualization of social and other scale-free networks.
\newblock \emph{IEEE transactions on visualization and computer graphics}, \emph{14}(6): 1285--1292.
\bibAnnoteFile{jia2008visualization}

\bibitem[{Karrer and Newman(2011)}]{karrer2011stochastic}
Karrer, Brian and Mark~EJ Newman. 2011.
\newblock Stochastic blockmodels and community structure in networks.
\newblock \emph{Physical review E}, \emph{83}(1): 016107.
\bibAnnoteFile{karrer2011stochastic}

\bibitem[{Kijima et~al.(2010)Kijima, Nishide, and Ohyama}]{Kijima2010}
Kijima, Masaaki, Katsumasa Nishide, and Atsuyuki Ohyama. 2010.
\newblock Economic models for the environmental Kuznets curve: A survey.
\newblock \emph{Journal of Economic Dynamics and Control}, \emph{34}(7): 1187--1201.
\bibAnnoteFile{Kijima2010}

\bibitem[{Koondhar et~al.(2021)Koondhar, Shahbaz, Memon, Ozturk, and Kong}]{Koondhar2021}
Koondhar, Mansoor~Ahmed, Muhammad Shahbaz, Kamran~Ali Memon, Ilhan Ozturk, and Rong Kong. 2021.
\newblock A visualization review analysis of the last two decades for environmental Kuznets curve “EKC” based on co-citation analysis theory and pathfinder network scaling algorithms.
\newblock \emph{Environmental Science and Pollution Research}, \emph{28}: 16690--16706.
\bibAnnoteFile{Koondhar2021}

\bibitem[{Kuznets(1955)}]{Kuznets1955}
Kuznets, Simon. 1955.
\newblock Economic {Growth} and {Income} {Inequality}.
\newblock \emph{The American Economic Review}, \emph{45}(1): 1--28.
\bibAnnoteFile{Kuznets1955}

\bibitem[{Leskovec and Faloutsos(2006)}]{leskovec2006sampling}
Leskovec, Jure and Christos Faloutsos. 2006.
\newblock Sampling from large graphs.
\newblock In \emph{Proceedings of the 12th ACM SIGKDD international conference on Knowledge discovery and data mining}. 631--636.
\bibAnnoteFile{leskovec2006sampling}

\bibitem[{Ma and Stern(2006)}]{Ma2006}
Ma, Chunbo and David~I. Stern. 2006.
\newblock Environmental and ecological economics: {A} citation analysis.
\newblock \emph{Ecological Economics}, \emph{58}(3): 491--506.
\bibAnnoteFile{Ma2006}

\bibitem[{Maiya and Berger-Wolf(2011)}]{maiya2011benefits}
Maiya, Arun~S and Tanya~Y Berger-Wolf. 2011.
\newblock Benefits of bias: Towards better characterization of network sampling.
\newblock In \emph{Proceedings of the 17th ACM SIGKDD international conference on Knowledge discovery and data mining}. 105--113.
\bibAnnoteFile{maiya2011benefits}

\bibitem[{Meadows et~al.(1972)Meadows, Meadows, Randers, and Behrens}]{Meadows1972}
Meadows, Donella~H., Dennis~L. Meadows, J\o{}rgen Randers, and William~W. Behrens. 1972.
\newblock \emph{The limits to growth: a report for the Club of Rome's project on the predicament of mankind}.
\newblock New-York: Universe Books.
\bibAnnoteFile{Meadows1972}

\bibitem[{Menezes and Roth(2019)}]{menezes2019semantic}
Menezes, Telmo and Camille Roth. 2019.
\newblock Semantic Hypergraphs.
\newblock \emph{arXiv}, \emph{1908.10784}.
\bibAnnoteFile{menezes2019semantic}

\bibitem[{Narayan and Narayan(2010)}]{Narayan2010}
Narayan, Paresh~Kumar and Seema Narayan. 2010.
\newblock Carbon dioxide emissions and economic growth: {Panel} data evidence from developing countries.
\newblock \emph{Energy Policy}, \emph{38}(1): 661--666.
\bibAnnoteFile{Narayan2010}

\bibitem[{Naveed et~al.(2022)Naveed, Ahmad, Aghdam, and Menegaki}]{Naveed2022}
Naveed, Amjad, Nisar Ahmad, Reza~FathollahZadeh Aghdam, and Angeliki~N Menegaki. 2022.
\newblock What have we learned from Environmental Kuznets Curve hypothesis? A citation-based systematic literature review and content analysis.
\newblock \emph{Energy Strategy Reviews}, \emph{44}: 100946.
\bibAnnoteFile{Naveed2022}

\bibitem[{Nordhaus(1992)}]{Nordhaus1992}
Nordhaus, William~D. 1992.
\newblock Lethal {Model} 2: {The} {Limits} to {Growth} {Revisited}.
\newblock \emph{Brookings Papers on Economic Activity}, \emph{1992}(2): 1--59.
\bibAnnoteFile{Nordhaus1992}

\bibitem[{Panayotou(1993)}]{Panayotou1993}
Panayotou, Theodore. 1993.
\newblock Empirical Tests and Policy Analysis of Environmental Degradation at Different Stages of Economic Development.
\newblock Geneva: International Labour Office.
\bibAnnoteFile{Panayotou1993}

\bibitem[{Panayotou(1997)}]{Panayotou1997}
Panayotou, Theodore. 1997.
\newblock Demystifying the environmental {Kuznets} curve: turning a black box into a policy tool.
\newblock \emph{Environment and Development Economics}, \emph{2}(4): 465--484.
\bibAnnoteFile{Panayotou1997}

\bibitem[{Peixoto(2018)}]{peixoto2018nonparametric}
Peixoto, Tiago~P. 2018.
\newblock Nonparametric weighted stochastic block models.
\newblock \emph{Physical Review E}, \emph{97}(1): 012306.
\bibAnnoteFile{peixoto2018nonparametric}

\bibitem[{Piketty(2014)}]{Piketty2014}
Piketty, Thomas. 2014.
\newblock \emph{Capital in the twenty-first century}.
\newblock Cambridge Massachusetts: The Belknap Press of Harvard University Press.
\bibAnnoteFile{Piketty2014}

\bibitem[{Ribeiro and Towsley(2010)}]{ribeiro2010estimating}
Ribeiro, Bruno and Don Towsley. 2010.
\newblock Estimating and sampling graphs with multidimensional random walks.
\newblock In \emph{Proceedings of the 10th ACM SIGCOMM conference on Internet measurement}. 390--403.
\bibAnnoteFile{ribeiro2010estimating}

\bibitem[{Ruggie(1975)}]{rugg-inte}
Ruggie, John~Gerard. 1975.
\newblock International Responses to Technology: Concepts and Trends.
\newblock \emph{International Organization}, \emph{29}(3): 557--583.
\bibAnnoteFile{rugg-inte}

\bibitem[{Sarkodie and Strezov(2019)}]{Sarkodie2019}
Sarkodie, Samuel~Asumadu and Vladimir Strezov. 2019.
\newblock A review on {Environmental} {Kuznets} {Curve} hypothesis using bibliometric and meta-analysis.
\newblock \emph{Science of The Total Environment}, \emph{649}: 128--145.
\bibAnnoteFile{Sarkodie2019}

\bibitem[{Schmid et~al.(2020)Schmid, Cowen, Robinson, Luo, Brise{\~n}o-Avena, and Sponaugle}]{schmid2020prey}
Schmid, Moritz~S, Robert~K Cowen, Kelly Robinson, Jessica~Y Luo, Christian Brise{\~n}o-Avena, and Su~Sponaugle. 2020.
\newblock Prey and predator overlap at the edge of a mesoscale eddy: Fine-scale, in-situ distributions to inform our understanding of oceanographic processes.
\newblock \emph{Scientific Reports}, \emph{10}(1): 921.
\bibAnnoteFile{schmid2020prey}

\bibitem[{Schmitt(1995)}]{schm:soci}
Schmitt, Frederik (ed.). 1995.
\newblock \emph{Socializing Epistemology: The Social Dimensions of Knowledge}.
\newblock Lanham, MD: Rowman \& Littlefield.
\bibAnnoteFile{schm:soci}

\bibitem[{Shafik and Bandyopadhyay(1992)}]{Shafik1992}
Shafik, Nemat and Sushenjit Bandyopadhyay. 1992.
\newblock Economic Growth and Environmental Quality: Time Series and Crosscountry Evidence. Background Paper for the World Development Report 1992.
\newblock Washington: The World Bank.
\bibAnnoteFile{Shafik1992}

\bibitem[{Shahbaz and Sinha(2019)}]{Shahbaz2019}
Shahbaz, Muhammad and Avik Sinha. 2019.
\newblock Environmental {Kuznets} curve for {CO2} emissions: a literature survey.
\newblock \emph{Journal of Economic Studies}, \emph{46}(1): 106--168.
\bibAnnoteFile{Shahbaz2019}

\bibitem[{Sterling et~al.(1995)Sterling, Rosenbaum, and Weinkam}]{Sterling1995}
Sterling, T.~D., W.~L. Rosenbaum, and J.~J. Weinkam. 1995.
\newblock Publication {Decisions} {Revisited}: {The} {Effect} of the {Outcome} of {Statistical} {Tests} on the {Decision} to {Publish} and {Vice} {Versa}.
\newblock \emph{The American Statistician}, \emph{49}(1): 108--112.
\bibAnnoteFile{Sterling1995}

\bibitem[{Sterling(1959)}]{Sterling1959}
Sterling, Theodore~D. 1959.
\newblock Publication {Decisions} and their {Possible} {Effects} on {Inferences} {Drawn} from {Tests} of {Significance}--or {Vice} {Versa}.
\newblock \emph{Journal of the American Statistical Association}, \emph{54}(285): 30--34.
\bibAnnoteFile{Sterling1959}

\bibitem[{Stern(2004)}]{Stern2004}
Stern, David~I. 2004.
\newblock The {Rise} and {Fall} of the {Environmental} {Kuznets} {Curve}.
\newblock \emph{World Development}, \emph{32}(8): 1419--1439.
\bibAnnoteFile{Stern2004}

\bibitem[{Stern(2010)}]{Stern2010}
Stern, David~I. 2010.
\newblock Between estimates of the emissions-income elasticity.
\newblock \emph{Ecological Economics}, \emph{69}(11): 2173--2182.
\bibAnnoteFile{Stern2010}

\bibitem[{Stern(2017)}]{Stern2017}
Stern, David~I. 2017.
\newblock The environmental {Kuznets} curve after 25 years.
\newblock \emph{Journal of Bioeconomics}, \emph{19}(1): 7--28.
\bibAnnoteFile{Stern2017}

\bibitem[{Stern and Common(2001)}]{Stern2001}
Stern, David~I. and Michael~S. Common. 2001.
\newblock Is {There} an {Environmental} {Kuznets} {Curve} for {Sulfur}?
\newblock \emph{Journal of Environmental Economics and Management}, \emph{41}(2): 162--178.
\bibAnnoteFile{Stern2001}

\bibitem[{Stern et~al.(1996)Stern, Common, and Barbier}]{Stern1996}
Stern, David~I., Michael~S. Common, and Edward~B. Barbier. 1996.
\newblock Economic growth and environmental degradation: {The} environmental {Kuznets} curve and sustainable development.
\newblock \emph{World Development}, \emph{24}(7): 1151--1160.
\bibAnnoteFile{Stern1996}

\bibitem[{Van~Ham and Wattenberg(2008)}]{van2008centrality}
Van~Ham, Frank and Martin Wattenberg. 2008.
\newblock Centrality based visualization of small world graphs.
\newblock \emph{Computer Graphics Forum}, \emph{27}(3): 975--982.
\bibAnnoteFile{van2008centrality}

\bibitem[{Voudigari et~al.(2016)Voudigari, Salamanos, Papageorgiou, and Yannakoudakis}]{voudigari2016rank}
Voudigari, Elli, Nikos Salamanos, Theodore Papageorgiou, and Emmanuel~J Yannakoudakis. 2016.
\newblock Rank degree: An efficient algorithm for graph sampling.
\newblock In \emph{2016 IEEE/ACM International Conference on Advances in Social Networks Analysis and Mining (ASONAM)}. IEEE, 120--129.
\bibAnnoteFile{voudigari2016rank}

\bibitem[{Wagner(2008)}]{Wagner2008}
Wagner, Martin. 2008.
\newblock The carbon {Kuznets} curve: {A} cloudy picture emitted by bad econometrics?
\newblock \emph{Resource and Energy Economics}, \emph{30}(3): 388--408.
\bibAnnoteFile{Wagner2008}

\bibitem[{Wagner(2015)}]{Wagner2015}
Wagner, Martin. 2015.
\newblock The {Environmental} {Kuznets} {Curve}, {Cointegration} and {Nonlinearity}.
\newblock \emph{Journal of Applied Econometrics}, \emph{30}(6): 948--967.
\bibAnnoteFile{Wagner2015}

\bibitem[{{World Bank}(1992)}]{WB1992}
{World Bank}. 1992.
\newblock World Development Report 1992. Development and the Environment.
\newblock New York: World Bank.
\bibAnnoteFile{WB1992}

\bibitem[{Zhao et~al.(2020)Zhao, Jiang, Qin, Xie, Wu, Liu, Zhou, Xia, Zhou et~al.}]{zhao2020preserving}
Zhao, Ying, Haojin Jiang, Yaqi Qin, Huixuan Xie, Yitao Wu, Shixia Liu, Zhiguang Zhou, Jiazhi Xia, Fangfang Zhou et~al. 2020.
\newblock Preserving minority structures in graph sampling.
\newblock \emph{IEEE Transactions on Visualization and Computer Graphics}, \emph{27}(2): 1698--1708.
\bibAnnoteFile{zhao2020preserving}

\end{thebibliography}

\appendix

\section{Semantic hypergraphs}\label{app-sh}
We provide here a brief theoretical and practical overview of SH, especially in the context of this article. Interested readers may find further details in \citep{menezes2019semantic}.

First, SH are made of (semantic) hyperedges, representing utterances as relations consisting of a predicate (annotated with \Tx{/P}) followed by an arbitrary number of participants (concepts are annotated with \Tx{/C}). The building blocks of SH are \emph{atoms}, which mostly correspond to words annotated with types (a few additional atoms are defined, for example to define connectors for compound nouns). The sentence ``Mary plays chess.'' translates to:
\QCx{(plays/P mary/C chess/C)}
Relations can be nested, for example in ``John says Mary plays chess.'':
\QCx{(says/P john/C (plays/P mary/C chess/C))}
Relations can be combined with conjunctions (annotated with \Tx{/J}), for example in ``John reads and Mary plays chess.'':
\QCx{(and/J (reads/P john/C) (plays/P mary/C chess/C))}
Relations connect concepts, but concepts themselves can be recursively constructed with other concepts with the help of builders (\Tx{/B}), such as in ``Mayor of Berlin'':
\QCx{(of/B mayor/C berlin/C)}

Finally, concepts, predicates and other hyperedges can be made more specific with modifiers (\Tx{/M}):
\QCx{(first/M (of/B mayor/C berlin/C))}

The above recursivity is an essential property of SHs in that they mirror the recursive scaffolding of natural language sentences:
\QCx{(plays/P (the/M (last/M (of/B mayor/C berlin/C))) (chess/C))}

\paragraph{Defining hypergraphic rules to detect EKC-related results.}
We used Graphbrain\footnote{https://graphbrain.net}, an open source library that implements a variety of functions related to SH, to convert all the sentences in article abstracts to hyperedges. Then, we took advantage of the structure and regularity imposed on natural language by the SH representation to define rules that detect the reporting of results either confirming or refuting EKCs.

We focus on relations with predicates that are likely to be related with reporting results, e.g.: ``found''.
\begin{equation}H=\text{\Tx{(p s a a' ...)}}\end{equation}
where \Tx{p} is of type predicate (\Tx{/P}), and the (optional) arguments are denoted as \Tx{s} for the potential subject of the relation, and \Tx{a}, \Tx{a'}, \hbox{etc.} for non-subject arguments. Graphbrain is particularly efficient at determining the semantic roles fulfilled by the sequence of arguments of a relation hyperedge \citep[we refer the interested reader to][for more details]{menezes2019semantic}.

We then iteratively refined a small set of rules to extract hyperedges denoting results related to the empirical confirmation or refutation of EKC. To this end, we used a subset of the corpus made of 500 randomly selected sentences, out of 23042 (2.17\%), akin to a training set in conventional machine learning. The main steps were as follows:
\itx{
\x We looked for the most commonly used predicates and extracted the ones usually associated with results: we explored 100 top predicates using the Graphbrain tool, which offers this functionality, and found 20 of them (shown in table~\ref{tab:atom-sets}) to more frequently correspond to claims about results. 

The first main rule is that \Tx{p} contains an atom belonging to this set of so-called claim predicates, either positive ($P^+$) or negative ($P^-$).

\begin{table}[t!]
\begin{center}
\begin{tabularx}{\linewidth}{>{\em}l@{\hspace{.7em}}>{\em}p{1.7cm}l}
\toprule
Set && Atoms \\
\midrule
\multirow{6}{*}{Claim predicates}&\multirow{5}{*}{positive ($P^+$)}& show/P, indicate/P, confirm/P, support/P, \\
                                 & & suggest/P, reveal/P, provide/P, validate/P, exist/P, \\
                                 & & demonstrate/P, verify/P, imply/P, illustrate/P, \\
                                 & & find/P, point/P, exhibit/P, establish/P, obtain/P, \\
                                 & & hold/P, follow/P \smallskip\\\cline{2-3}
&negative~($P^-$) & reject/P, challenge/P, fail/P \\
\midrule
EKC concepts && kuznets/C, ekc/C, turning/C \\
Curve concepts && curve/C, shape/C, shaped/C
\\
\multicolumn{2}{l}{\em Negative modifiers}& not/M, n't/M, no/M, little/M, poor/M \\
Result concepts &($R$)
 & result/C, finding/C, test/C, evidence/C, support/C \\

\bottomrule
\end{tabularx}
\caption{The various atom sets used in claim detection and classification rules.}
\label{tab:atom-sets}
\end{center}
\end{table}

\x Many results are not directly about EKC, so we limited the rules to explicit references to EKC. We also included rules to capture the notions of ``U-curves'' and ``N-curves'' which, in the corpus, generally refer implicitly to EKC.

In practice, we thus defined a helper rule $\mathcal{E}$ as follows:
$\mathcal{E}(H)$ true if $\exists \Tx{a}\in H$ such that,
	\itx{
	\x either $\Tx{a}\in \{\Tx{kuznets/C}, \Tx{ekc/C}, \Tx{turning/C}\}$
	\x or $\exists \Tx{c}\in \{\Tx{curve/C}, \Tx{shape/C}, \Tx{shaped/C}
	\}$,\\such that $\Tx{a}=\text{\Tx{(u/C c)}} \text{\; or \;} \Tx{a}=\text{\Tx{(n/C c)}}$
	}

\x However, if the reference to EKC is in the subject, it is not a result (``EKCs reveal~(...)'' \hbox{vs.} ``We reveal the presence of EKC''), and we include a rule where $\mathcal{E}(\Tx{s})$ must be false.
}

The following expression formalizes the above rules for identifying a result claim about EKCs, as a logical expression (whereby `$\wedge$' and `$\neg$' denote respectively the `AND' and `NOT' logical operators):
\begin{equation}
    \mathcal{C}(H) = \left(\Tx{p} \cap (P^+\cup P^-)\ne\varnothing\right) \wedge \left(\mathcal{E}(H) \wedge \neg\mathcal{E}(\Tx{(s)})\right)
\end{equation}

\paragraph{Positive \hbox{vs.} negative results.}
To determine if a result claim is positive, negative or unknown in regard to EKC validity, we further take advantage of SH structure. We add a helper rule $\mathcal{N}$ to detect negative claims \hbox{i.e.,} cases where the predicate is negated (e.g.: ``Could not find evidence of...''), or the concept referring to the result is negated (e.g.: ``We found no evidence of...'') , or the concept referring to EKC is negated (e.g.: ``No EKC was found...''). Formally, $\mathcal{N(H)}$ is true if $H$ contains:
\itx{
\x either, both:
 \itx{
 \x a negation in its predicate \Tx{p}: \\
 $\Tx{p}\cap \{\Tx{not/M}, \Tx{n't/M}, \Tx{no/M}, \Tx{little/M}, \Tx{poor/M}\} \ne\varnothing$
\x and a claim predicate or a result concept in one of its arguments:\\
$\exists \Tx{a}\in H,
	\Tx{a}\cap (P\cup R)\ne\varnothing
	\text{\; or \;}
	\mathcal{E}(\Tx{a})$
	
	\noindent which further relies on a set $R$ of atoms of so-called ``result concepts'': $$R=\{\text{\Tx{result/C, finding/C, test/C, evidence/C, support/C}}\}$$ 
\noindent This set has been constructed by examining the most frequent atoms directly connected to negations (among the top 100 such atoms), similarly to how $P$ has been built.
}
\x or, recursively, an element of $H$ (possibly a hyperedge itself) for which $N$ is true:
$\exists h\in H, \mathcal{N}(h)$ is true.
}

Finally, table~\ref{tab:pos-neg-claims} shows how a result claim is classified. Notice that we consider a positive finding of an N-curve (\hbox{i.e.,} $\exists \Tx{a}\in H, \Tx{a}=\text{\Tx{(n/C c)}}$), as a refutation of EKC. We do not consider double negations, as we found them to increase complexity while having no impact on accuracy.

\begin{table}[h!]
\newcommand{\true}{\multirow{2}{*}{True}}
\newcommand{\false}{\multirow{2}{*}{False}}
\begin{center}\setlength{\tabcolsep}{1em}
\begin{tabular}{ccc>{\em}c}\toprule
predicate $p$&\multirow{2}{*}{$\mathcal{N}(H)$}&$H$ contains& EKC validation\\
based on&&an N-curve&claim type\\
\toprule
\multirow{4}{*}{$P^+$}  & \true & True & X \\
  &  & False & negative result \\\cline{2-4}
  & \false & True & negative result \\
  &  & False & positive result\\
  \midrule
\multirow{4}{*}{$P^-$}  & \true &  & \multirow{2}{*}X \\
  & & &  \\\cline{2-4}
  & \false & True & X \\
  &  & False & negative result\\
\bottomrule
\end{tabular}
\caption{Classification of a claim $H$ as a positive or negative result, or unknown (X) based on the elements of $H$.}
\label{tab:pos-neg-claims}
\end{center}
\end{table}

\end{document}